\documentclass[11pt]{article}
\RequirePackage[OT1]{fontenc}
\RequirePackage{amsthm,amsmath}
\RequirePackage{natbib}
\RequirePackage[colorlinks,citecolor=blue,urlcolor=blue]{hyperref}
\usepackage{amsmath}
\usepackage{bbm}
\usepackage{bm}
\usepackage{amssymb}
\usepackage{extarrows}
\usepackage{framed}
\usepackage{rotating}
\usepackage{graphicx}
\usepackage{graphics}
\usepackage{hyperref}
\usepackage{comment}
\usepackage{algorithm2e}
\usepackage{float}
\usepackage{bm}
\usepackage{subcaption}
\usepackage[symbol]{footmisc}
\usepackage{lineno}
\usepackage{authblk}
\usepackage{geometry}
\makeatletter
\def\blfootnote{\xdef\@thefnmark{}\@footnotetext}
\makeatother

\newcommand*\samethanks[1][\value{footnote}]{\footnotemark[#1]}

\begin{document}
\title{Late 19th-Century Navigational Uncertainties and Their Influence on Sea Surface Temperature Estimates}
\author[1]{Chenguang Dai\thanks{These authors contributed equally to this work.}}
\author[2]{Duo Chan\samethanks}
\author[2]{Peter Huybers}
\author[1]{Natesh Pillai}

\affil[1]{Department of Statistics, Harvard University}
\affil[2]{Department of Earth and Planetary Sciences, Harvard University}

\maketitle

\begin{abstract}

Accurate estimates of historical changes in sea surface temperatures (SSTs) and their uncertainties are important for documenting and understanding historical changes in climate.  A source of uncertainty that has not previously been quantified in historical SST estimates stems from position errors.  A Bayesian inference framework is proposed for quantifying errors in reported positions and their implications on SST estimates.  The analysis framework is applied to data from the International Comprehensive Ocean-Atmosphere Data Set (ICOADS3.0) in 1885, a time when astronomical and chronometer estimation of position was common, but predating the use of radio signals.  Focus is upon a subset of 943 ship tracks from ICOADS3.0 that report their position every two hours to a precision of $0.01^\circ$ longitude and latitude.  These data are interpreted as positions determined by dead reckoning that are periodically updated by celestial correction techniques.  
The posterior medians of uncertainties in celestial correction are 33.1 km ($0.30^\circ$ on the equator) in longitude and 24.4 km ($0.22^\circ$) in latitude, respectively. 
Celestial navigation uncertainties being smaller in latitude than longitude is qualitatively consistent with the relative difficulty of obtaining astronomical estimates.  
The posterior medians  for two-hourly dead reckoning uncertainties are $19.2\%$ for ship speed and $13.2^\circ$ for ship heading, leading to random position uncertainties with median $0.18^\circ$(20 km on the equator) in longitude and $0.15^\circ$(17 km) in latitude.  
Reported ship tracks also contain systematic position uncertainties relating to precursor dead-reckoning positions not being updated after obtaining celestial position estimates, indicating that more accurate positions can be provided for SST observations.  Finally, we translate position errors into SST uncertainties by sampling an ensemble of SSTs from the Multi-scale Ultra-high resolution Sea Surface Temperature (MURSST) data set.  Evolving technology for determining ship position, heterogeneous reporting and archiving of position information, and seasonal and spatial changes in navigational uncertainty and SST gradients together imply that accounting for positional error in SST estimates over the span of the instrumental record will require substantial additional effort.

\end{abstract}

%%%%%%%%%%%%%%%%%%%%%%%%%%%%%%%%%%%%%%%%%%%%%%%%%%%%%%%%%%%%%%%%%%%%%%%%%%%%%
\section{Introduction}

Accurate estimates of past sea surface temperatures (SSTs) are important for assessing historical climate states \citep{morice2012quantifying}, detecting and attributing changes in climate \citep{chan2015attributing}, and computing climate sensitivity \citep{gregory2002observationally}.  SST datasets are also used as boundary conditions to run general circulation models \citep{folland2005assessing,sobel2007longitude}, and are assimilated as part of generating atmospheric reanalysis data sets \citep{dee2011era}.  SST datasets are, however, known to have substantial errors \citep{kent2017call}, especially prior to the systematic satellite, drifters, and moored buoy temperatures that became routinely available in the 1980s \citep{kennedy2011reassessing}.   For SST, quantified errors include those associated with random errors of individual measurements \citep{kent2006toward,ingleby2010factors}, systematic errors associated with different measurement methods \citep{kennedy2011reassessingB,huang2017extended}, offsets amongst different groups of observers \citep{chan2019correcting} as well as those associated with individual ships \citep{kennedy2012using}.  Another important source of uncertainty involves mapping noisy and often sparse observations to infill unobserved locations \citep{kennedy2014review}.

Despite quantification of many contributors to SST uncertainties, we are unaware of previous studies having quantitatively assessed navigational uncertainties associated with historical ocean observations.  That is, errors in position associated with incorrectly recording or transcribing locations have been recognized \citep{woodruff1998coads}, as have errors introduced by rounding of positions \citep{kent1999statistical}, but the magnitude of navigational uncertainties prior to the widespread deployment of radio navigation in the 1930s \citep{fried1977comparative} appears not to have been quantified.  Prior to radio navigation, ship position in the open ocean was mainly estimated by dead reckoning and celestial techniques \citep{bowditch1906american}.  Dead reckoning involves updating ship position using estimates of heading and distance.  Celestial navigation involves estimating latitude from the zenith angle associated with various celestial bodies, including the sun, moon, and stars.  Longitude may be inferred using a chronometer method whereby the difference between a local apparent time and the time at some known longitude are determined from a clock carried onboard or some other method, such as the phase of Jupiter's moons.  Dead reckoning can potentially introduce both systematic and random uncertainties, whereas celestial correction is assumed free of systematic uncertainties.

Position errors have implications on the accuracy of mapped SSTs.  For example, if SST measurements are binned into gridboxes, misspecification of the appropriate box will influence the mean and higher-order moments \citep{director2015connecting}.  \cite{cervone2015gaussian} have shown that incorporating position uncertainties when averaging land-station data within gridboxes, which have typically been assumed to reside at the center of the gridbox, is important for valid inference of land surface temperatures, and we expect that the the additional uncertainties over the sea associated with ship positions are no less important.  In the following we propose a Bayesian model to quantify position errors for various ship tracks through estimating navigational uncertainties in dead reckoning and celestial correction.  We then translate position errors into SST uncertainties by sampling high-resolution SSTs using posterior samples of ship positions.

\section{Data description}{\label{sec:data}}

The ship data used in this study are from the International Comprehensive Ocean-Atmosphere Data Set (ICOADS3.0) \citep{freeman2016icoads}, which is the most comprehensive available historical dataset of ship-based measurements from the eighteenth century to the present. We use data in 1885 to demonstrate a Bayesian framework for estimating position errors associated with historical ship tracks. Individual ship tracks are identified using ICOADS identification (ID) information, and tracks with missing or non-unique IDs are excluded. Ship tracks traversing across open ocean are separated into shorter segments whenever they are close to islands.  Only ship tracks that have their positions reported at a resolution of every four hours or better and to a precision of better than one degree longitude and latitude are retained.  These high-resolution ship tracks are focussed on because it is otherwise difficult to identify positions errors and to distinguish between contributions from dead-reckoning and celestial navigation.

\begin{figure*}
\centering
\includegraphics[width=30pc]{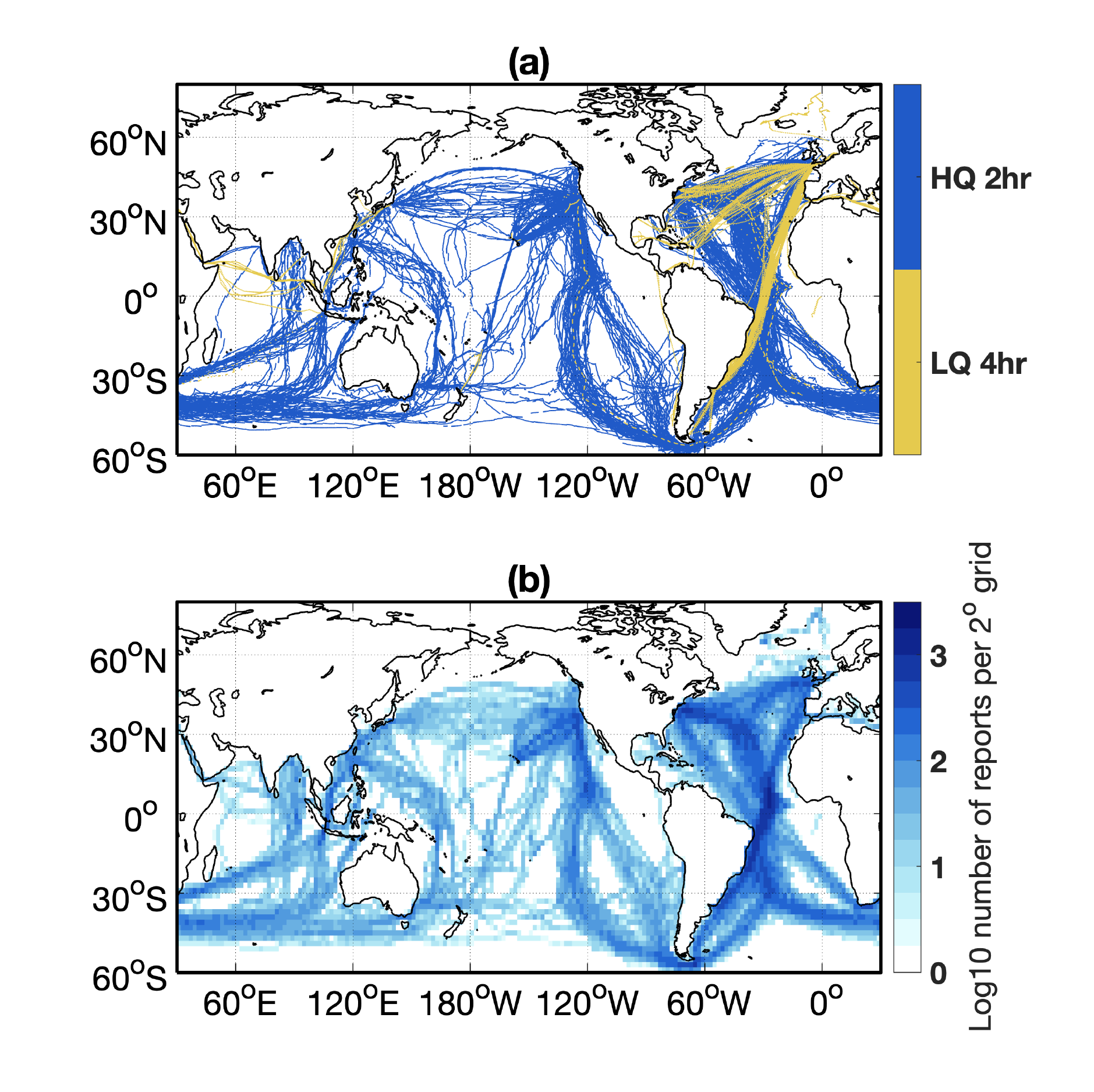}
\caption{Ship tracks used in this study. (a) Tracks are divided into high-quality, two-hourly ship tracks (HQ2) and low-quality, four-hourly ship tracks (LQ4). (b) Density of ship position reports at 2$^\circ$ resolution.}
\label{fig:AllTracks}
\end{figure*}

\begin{figure*}
\centering
\includegraphics[width=30pc]{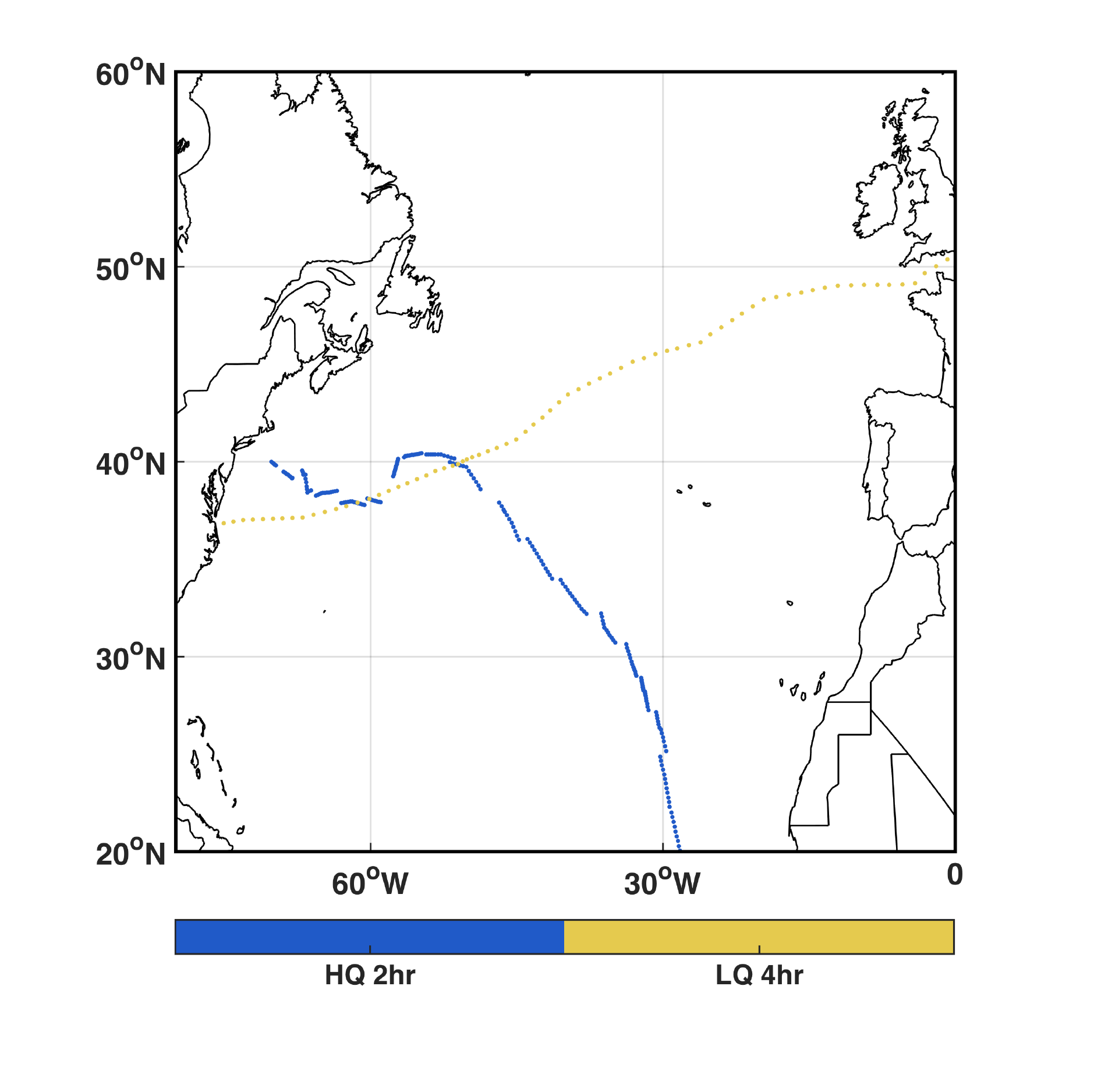}
\caption{Two types of ship tracks. The blue track is the high-quality 2-hourly ship track No.30, and the yellow track is the low-quality 4-hourly ship track No.108. Jumps in position are seen in the 2-hourly track, whereas the 4-hourly track appears overly smooth.}
\label{fig:ThreeTypesRaw}
\end{figure*}

The highest resolution data comes from the U.S.~Marine Meteorological Journals Collection, which was a program sponsored by the U.S.~Navy's Hydrographic Office that enlisted the help of commercial vessels in compiling meteorological data.  Reports are primarily, albeit not exclusively, from U.S.~vessels and are provided every two hours at a resolution of 0.01$^\circ$ longitude and latitude.  In total there are 1,341 of these two-hourly ship tracks.  943 of these tracks are characterized by stable velocities that are episodically punctuated by jumps in position (see Figure \ref{fig:ThreeTypesRaw} and Table \ref{table:comparisonEmspJumpDist}).  Jumps in otherwise smooth ship tracks typically occur at midnight and are consistent with navigation using dead reckoning that is updated by a celestial positioning technique \citep{bowditch1906american}.  Ship tracks generally follow well-established trade routes that tend to be meridional in the tropics and zonal in the mid-latitudes, with the highest data density in the Atlantic, the Eastern Pacific, and the Southern Indian Ocean (see Figure \ref{fig:AllTracks}).

The remaining 398 two-hourly tracks show static positions followed by jumps averaging $84.6$ km in 2 hours, which is unphysical for a ship under sail.  We are unaware of metadata indicating how these positions were prescribed, and exclude these tracks from our present analysis.

\begin{table*}
\centering
\caption{Empirical speed and the jumping distance.}
\begin{tabular}{|c|ccc|ccc|}
\hline
& \multicolumn{3}{c}{Empirical speed (km/hr)} & \multicolumn{3}{|c|}{Jumping distance (km)} \\
\hline
Quantiles &\hspace{0.15cm} 25\% \hspace{0.15cm} & 50\% & 75\% &\hspace{0.15cm} 25\%\hspace{0.15cm} & 50\% & 75\% \\
\hline
HQ2 & 6.4 & 10.4 & 14.6 & 15.0 & 22.8 & 37.0 \\           
LQ4 & 15.8 & 18.3 & 19.6 & - - & - - & - - \\           
\hline
\end{tabular}
\label{table:comparisonEmspJumpDist}
\end{table*}

There also exists a separate collection of 576 ship tracks that report position every four hours to a precision of 0.01$^\circ$ longitude and latitude.  These four-hourly tracks, referred to as LQ4 tracks, primarily track zonally between Europe and North America, and meridionally between Europe and South America.  Unlike HQ2 tracks, LQ4 tracks appear overly smooth, showing no discontinuities as would be expected from celestial navigational updates. We assume that the position reports of these tracks have been manually interpolated, implying that they contain less useful information for purposes of inferring navigational uncertainties.  
Although LQ4 tracks are unreliable for inferring underlying navigational uncertainties, these more smoothly-varying tracks are more generally representative of position data available in ICOADS, and we develop a methodology for exploring their position uncertainties that leverages results obtained from HQ2 tracks.

It is necessary to define when celestial updates occur for HQ2 tracks.  Celestial navigational updates are presumed to occur when the ship track jumps. Using the speed and heading from neighboring ship positions, we predict the next position, and a jump is identified when the predicted and reported positions differ by at least 7 km in either longitude or latitude.  7 km is chosen on account of its being the 80th percentile of latitudinal differences between predicted and reported ship positions.  Note that longitude and latitude are treated independently because their respective methods of celestial positioning are distinct.  When several jumps are identified in a single day, only the largest jump is selected.

%%%%%%%%%%%%%%%%%%%%%%%%%%%%%%%%%%%%%%%%%%%%%%%%%%%%%%%%%%%%%%%%%%%%%%%%%%%%%
\section{Bayesian model}{\label{sec:bayesmodel}}

The proposed Bayesian model for estimating position errors contains three stages. First, position errors, uncertainties in ship speed and heading are inferred for each HQ2 track using a state-space time series model.  Second, navigational uncertainties are synthesized across different HQ2 tracks using a Bayesian hierarchical model.  Finally, uncertainties are modeled for LQ4 data using a forward navigation model based upon results obtained from HQ2 data.  Stages two and three utilize the posterior samples obtained from prior stages. 
All the models described below are fitted using RStan \citep{rstan}.

We note that ideally, these stages would be integrated into a single, inclusive, hierarchical Bayesian model. However, as we will see, for each ship track, the number of parameters in the model is approximately two times the length of the track. Based on our experiences, it can take up to six hours using RStan to fit a single ship track with approximate 400 reported positions. Therefore, given the amount of data as well as the model complexity, we proceed with the multi-stage approach mentioned above to bypass the computational difficulty of implementing a full-Bayesian procedure. 

\subsection{State-space model for HQ2 tracks}
\label{section:learnable 2-hourly}

The proposed model utilizes the reported HQ2 ship positions to empirically calculate ship speed and heading at two-hourly time steps.
Ship position and heading are in radians and speed is in km/hr.  Let $\left(q_t^x, q_t^y\right)$ be the displacements (km) that the ship travels from the starting position to the \textit{reported} position at time step $t$ using dead reckoning. Correspondingly, let $\left(p_t^x, p_t^y\right)$ be the displacements from the starting position to the \textit{true} position. $q_t^x, p_t^x$ ($q_t^y, p_t^y$) are positive if the current ship position is to the east (north) of the starting point. $q_t^x, q_t^y$ are calculated as accumulated sums following Equation \eqref{eq:qt}, where $\phi$ denotes the reported longitude, $\psi$ denotes the reported latitude, $r_a$ denotes the radius of earth, and the subscripts denote the time step. The cosine term accounts for changes in distance-longitude scaling with latitude. Definitions for parameters used in the model are listed in Table \ref{tab:parameters}.
\begin{equation}
q_t^x  = \sum_{i = 1}^t r_a \left(\phi_i - \phi_{i - 1}\right)\cos\left(\frac{\psi_i + \psi_{i - 1}}{2}\right),\ \ \
q_t^y = \sum_{i = 1}^t r_a (\psi_{i} - \psi_{i - 1}).
\label{eq:qt}
\end{equation}

\begin{table*}
\begin{minipage}{\textwidth}
\begin{center}
\caption{Definition of parameters}
\label{tab:parameters}
\begin{tabular}{l l}
\hline
Parameter\blfootnote{The subscripts or superscripts $x, y, s, \theta$ refer to the longitudinal direction, the latitudinal direction, ship speed and ship heading, respectively.} & Definition \\         \hline
$\left(\phi_t, \psi_t\right)$ & Reported ship position in longitude and latitude   \\
$\left(q_t^x, q_t^y\right)$ & Displacement from the starting position to the reported position\\
$\left(p_t^x, p_t^y\right)$ & Displacement from the starting position to the true position \\
$s_t$, $\hat{s}_t$, $\mu_s$ & True ship speed, empirical ship speed, mean of ship speed\\
$\theta_t$, $\hat{\theta}_t$, $\beta$ & True ship heading, empirical ship heading and its systematic bias \\
$\sigma_s, \sigma_\theta$ & Evolutionary uncertainties in ship speed and heading \\
$\tau_x, \tau_y$ & Track-based uncertainties in celestial correction\\
$\tau_s, \tau_\theta$ & Track-based uncertainties in dead reckoning \\
$\mu_{\tau_x}, \mu_{\tau_y}$ & Population median of uncertainties in celestial correction\\
$\mu_{\tau_s}, \mu_{\tau_\theta}$ & Population median of uncertainties in dead reckoning \\
\hline
\end{tabular}
\end{center}
\end{minipage}
\end{table*}

\subsubsection{Transition model}
For ship speed $s_t$, we assume a customary centered AR model described as below,
\begin{equation}
\label{eq:transition-sp}
s_t  =  \mu_s + \alpha_s(s_{t - 1} - \mu_s) + \epsilon^s_t,
\end{equation}
in which $\alpha_s \in (0, 1)$ denotes the drift parameter, and $\mu_s$ denotes the unknown \textit{track-based population} mean of $s_t$. 
Conditioning on $\alpha_s$, $\mu_s$ and $s_{t - 1}$, we assume that $\epsilon_t^s$ follows a truncated normal distribution with mean 0, variance $\sigma_s^2$, and lower truncation at $-\mu_s - \alpha_s(s_{t - 1} - \mu_s)$, so as to guarantee that $s_t$ is non-negative.

For ship heading $\theta_t$, we assume a simple random walk model. That is,
\begin{equation}
\label{eq:transition-th}
\theta_t = \theta_{t - 1} + \epsilon_t^\theta,\ \ \ \ \ \ \epsilon_t^\theta \sim N(0, \sigma_\theta^2).
\end{equation}
Both $\epsilon^s_t$ and $\epsilon^\theta_t$ are assumed to be independent across time.
The transition model is essentially parametrized in terms of $s_t$ and $\theta_t$, while the aggregated displacements $p_t^x$ and $p_t^y$ are deterministic functions of $s_t$ and $\theta_t$. That is,
\begin{equation}
p_t^x  = p_{t - 1}^x + 2s_t\cos\theta_t,\ \ \ \ \ \ p_t^y  = p_{t - 1}^y + 2s_t\sin\theta_t,
\end{equation}
which captures the physical navigation process by using ship speed and ship heading to project its next position (the same as dead reckoning navigation). 
We multiply $s_t$ by 2 in the model because we are modeling 2-hourly ship tracks, whereas $s_t$ is in unit of km/hr.

\subsubsection{Observation model}
Suppose there are in total $m$ celestial updates along the ship track. Let $\mathcal{C} = \{t_1, \cdots, t_m\}$ be the set of the corresponding time steps. 
\begin{figure}[h]
\centering
\includegraphics[width=20pc]{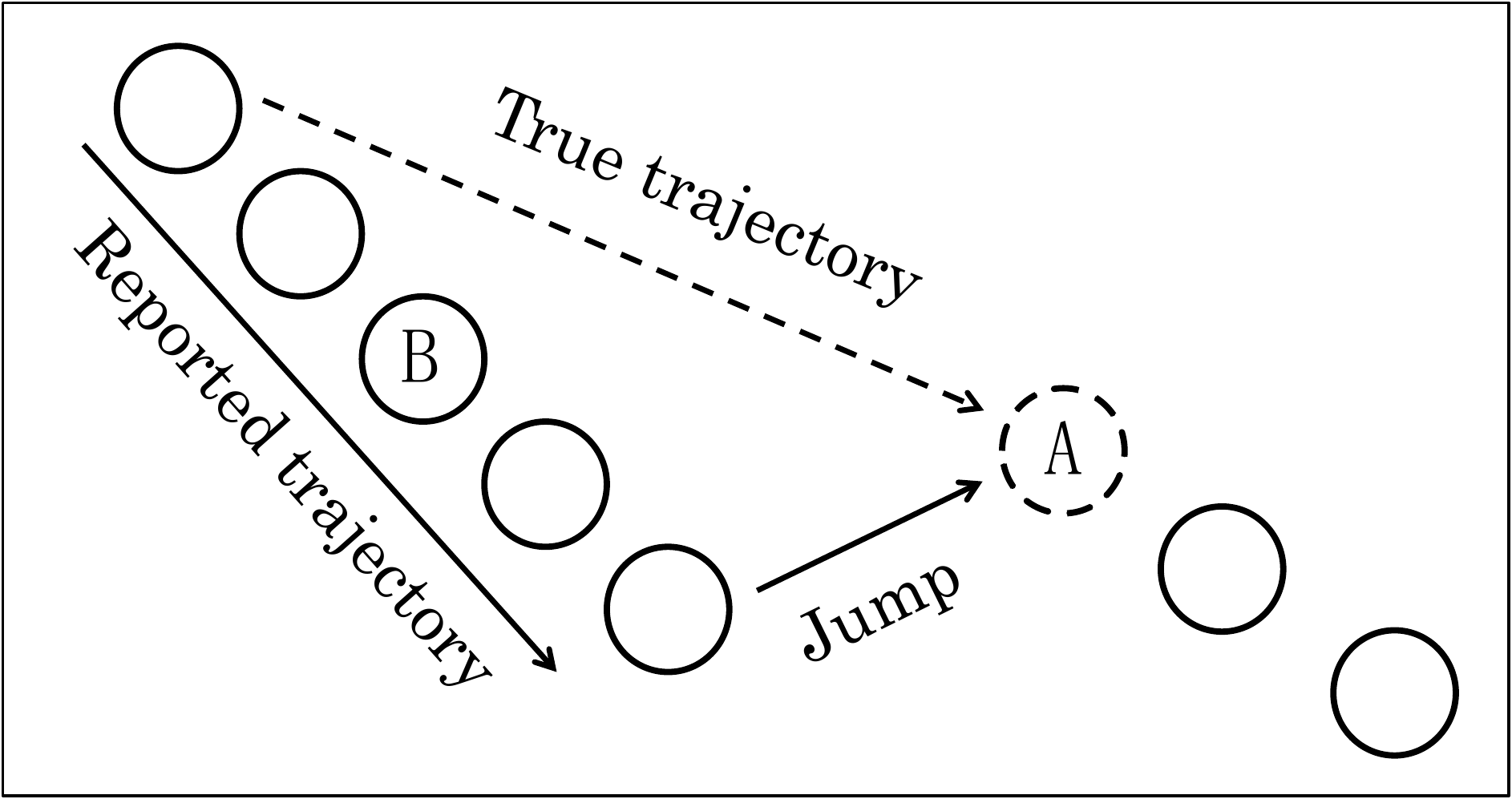}
\caption{A cartoon illustration of a celestial update along HQ2 tracks.  Reported ship positions are represented by circles, with the dashed circle representing a celestial correction.   The reported ship trajectory is indicated by the solid arrow, but it is systematically biased relative to the true ship trajectory indicated by the dashed arrow.}
\label{fig:illustration}
\end{figure}

For $t \not\in \mathcal{C}$ (e.g., position B in Figure \ref{fig:illustration}), we note that the reported ship trajectory is systematically biased, thus the ship position contains accumulated position errors.  Therefore, rather than basing the model on position information, empirical ship speed $\hat{s}_t$ and ship heading $\hat{\theta}_t$ are incorporated into the observation model, which are not expected to have persistence insomuch as dead-reckoning is utilized for navigation. Mathematically, for $k\in[1:(m - 1)]$ and $t \in (t_k, t_{k + 1})$, we assume
\begin{equation}
\label{eq:observation-dead-reckoning}
\hat{s}_t \sim N(s_t, (\tau_s s_t)^2), \ \ \ \ \hat{\theta}_t \sim N(\theta_t + \beta_k, \tau_\theta^2), \ \ \ \ 
\end{equation}
in which $\tau_s$ and $\tau_\theta$ denote the track-based uncertainties in dead reckoning. 
It is worthwhile to note that in early navigations, ship speed was usually measured by a chip log that measures the length of a rope released as a ship travels, or by a patent log that counts the numbers of turns a turbine rotates when water passes though \citep{bowditch1906american}. Since both methods tend to have larger measurement errors at higher ship speed, we model the uncertainty in terms of the relative ship speed instead of the absolute ship speed.
Besides, we introduce $\beta_k$ to model the systematic bias in ship heading between two celestial updates at time step $t_k$ and $t_{k + 1}$. The prior on $\beta_k$ is assumed to be $\text{Unif}(-\pi, \pi]$.

For $t \in \mathcal{C}$ (e.g., position A in Figure \ref{fig:illustration}), empirical ship speed and heading are ignored as they are very inaccurate because of the large jumps (see Table \ref{table:comparisonEmspJumpDist}), whereas the reported ship positions are assumed to contain only celestial observational errors. Thus, for $k\in[1:(m - 1)]$ and $t = t_k$, we assume
\begin{equation}
\label{eq:observation-celestial}
q_t^x \sim N(p_t^x, \left({\tau_x}\cos\psi_t\right)^2),\ \ \ \ \ q_t^y \sim N(p_t^y, {\tau_y}^2),
\end{equation}
in which $q_t^x$ and $q_t^y$ denote the observed aggregated displacements. $\psi_t$ denotes the reported latitude. $\tau_x$ and $\tau_y$ denote the track-based uncertainties in celestial correction in the longitudinal and latitudinal direction, respectively.

\subsection{Synthesizing information across different HQ2 tracks}
\label{section:BHM}
The state-space model discussed in Section \ref{section:learnable 2-hourly} describes a single HQ2 track.  We assume consistent levels of accuracy in celestial correction and dead reckoning over HQ2 tracks, permitting for borrowing information on the uncertainty parameters across different tracks.  Let $\{\tau_x^{(j)}, \tau_y^{(j)}, \tau_s^{(j)}, \tau_\theta^{(j)}\}$ be the uncertainty parameters associated with HQ2 track $j$. We assume a hierarchical structure on the uncertainty parameters,
\begin{equation}
\label{eq:idealBHM}
\begin{aligned}
& \log\tau_x^{(j)} \sim N(\log \mu_{\tau_x}, \gamma_{\tau_x}^2), \ \ \ \log\tau_y^{(j)} \sim N(\log\mu_{\tau_y}, \gamma_{\tau_y}^2).
\end{aligned}
\end{equation}
$\log\mu_{\tau_x}$ and $\log\mu_{\tau_y}$ denote the population medians of $\log\tau_x^{(j)}$ and $\log\tau_y^{(j)}$, respectively. Since the log-transformation is monotone, $\mu_{\tau_x}$ and $\mu_{\tau_x}$ are the population medians of $\tau_x^{(j)}$ and $\tau_y^{(j)}$, respectively.
$\tau_s^{(j)}, \tau_\theta^{(j)}$ are modeled in the same fashion.
Conditioning on the population-level parameters, track-based navigational parameters are assumed to be independent. 

To remedy the computational challenge, we consider a second-stage model, where we synthesize the uncertainty information across tracks based on the posterior samples obtained by fitting each HQ2 track. The variability of the navigational parameters across different tracks are summarized in Figure \ref{fig:variability} in the Appendix.

Given $n$ posterior samples $\{\tau_{x}^{(j, i)}, \tau_{y}^{(j, i)}\}_{i \in [1:n]}$ of HQ2 track $j$, we assume a hierarchical model,
\begin{equation}
\label{eq:BHM}
\begin{aligned}
\log\tau_{x}^{(j,i)} & \sim N\left(\log\mu^{(j)}_{\tau_x}, {\eta^{(j)}_{\tau_x}}^2\right),\ \ \ \log\mu^{(j)}_{\tau_x} \sim N(\log\mu_{\tau_x}, \gamma_{\tau_x}^2), \\
\log\tau_{y}^{(j,i)} & \sim N\left(\log\mu^{(j)}_{\tau_y}, {\eta^{(j)}_{\tau_y}}^2\right),\ \ \ \log\mu^{(j)}_{\tau_y} \sim N(\log\mu_{\tau_y}, \gamma_{\tau_y}^2),
\end{aligned}
\end{equation}
in which $\mu^{(j)}_{\tau_x}, \mu^{(j)}_{\tau_y}$ and $\eta^{(j)}_{\tau_x}, \eta^{(j)}_{\tau_y}$ denote the track-level medians and standard deviations for HQ2 track $j$. Conditioning on track-level parameters $\mu^{(j)}_{\tau_x}, \eta^{(j)}_{\tau_x}$ and $\mu^{(j)}_{\tau_y}, \eta^{(j)}_{\tau_y}$, we assume $\tau_{x}^{(j,i)}$ and $\tau_{y}^{(j,i)}$ are independent across the sample index $i$. Conditioning on the population-level parameters $\mu_{\tau_x}, \gamma_{\tau_x}$ and $\mu_{\tau_y}, \gamma_{\tau_y}$, we assume $\mu^{(j)}_{\tau_x}, \eta^{(j)}_{\tau_x}$ and $\mu^{(j)}_{\tau_y}, \eta^{(j)}_{\tau_y}$ are independent across the track index $j$. $\{\tau_{s}^{(j,i)}, \tau_{\theta}^{(j,i)}\}$ are modeled in the same fashion.

\subsection{Forward navigation model for LQ4 tracks}{\label{subsec:forward}}
The forward navigation model proposed in this section aims at representing dead reckoning and celestial correction contributions to position errors for LQ4 tracks.  
We assume that the population-level uncertainties for ship speed and heading in LQ4 tracks are consistent with the population-level uncertainties in HQ2 tracks.  Speed and heading can then be represented as,
\begin{equation}
\hat{s}_t = s_t(1 + e_t^s),\ \ \
\hat{\theta}_t = \theta_t + e_t^\theta,
\end{equation}
where $e_t^s, e_t^\theta$ are assumed to be i.i.d. over time. We impose the following hierarchical priors on $e_t^s, e_t^\theta$,
\begin{equation}
\label{eq:ideal-forward-model}
\begin{aligned}
e_t^s & \sim N(0, \tau_s^2),\ \ \ \log\tau_s \sim N(\log\hat{\mu}_{\tau_s}, \hat{\gamma}_{\tau_s}^2),\\
e_t^\theta & \sim N(0, \tau_\theta^2),\ \ \ \log\tau_\theta \sim N(\log\hat{\mu}_{\tau_\theta}, \hat{\gamma}_{\tau_\theta}^2),
\end{aligned}
\end{equation}
in which $\hat{\mu}_{\tau_s}$, $\hat{\mu}_{\tau_\theta}$, $\hat{\gamma}_{\tau_s}$ and $\hat{\gamma}_{\tau_\theta}$ are the empirical posterior means of $\mu_{\tau_s}$, $\mu_{\tau_\theta}$, $\gamma_{\tau_s}$ and $\gamma_{\tau_\theta}$, respectively, obtained by fitting the hierarchical model discussed in Section \ref{section:BHM}. The uncertainties in celestial correction are incorporated as in Equation \eqref{eq:observation-celestial}, and the priors on $\tau_x$ and $\tau_y$ are specified similarly as in Equation \eqref{eq:ideal-forward-model}.

For celestial navigation we incorporate a term representing the probability that a celestial observation is employed.
In order to illustrate the sensitivity of our results to whether celestial observations are taken --- and because we have no direct evidence of whether celestial observations are employed on any given day --- we provide results assuming observations are taken every night ($p=1$), half of the time ($p=0.5$), and never ($p=0$).

%%%%%%%%%%%%%%%%%%%%%%%%%%%%%%%%%%%%%%%%%%%%%%%%%%%%%%%%%%%%%%%%%%%%%%%%%%%%%
\section{Results}{\label{sec:results}}

\subsection{High-quality two-hourly ship tracks}
\begin{figure*}
\centering
\includegraphics[width=30pc]{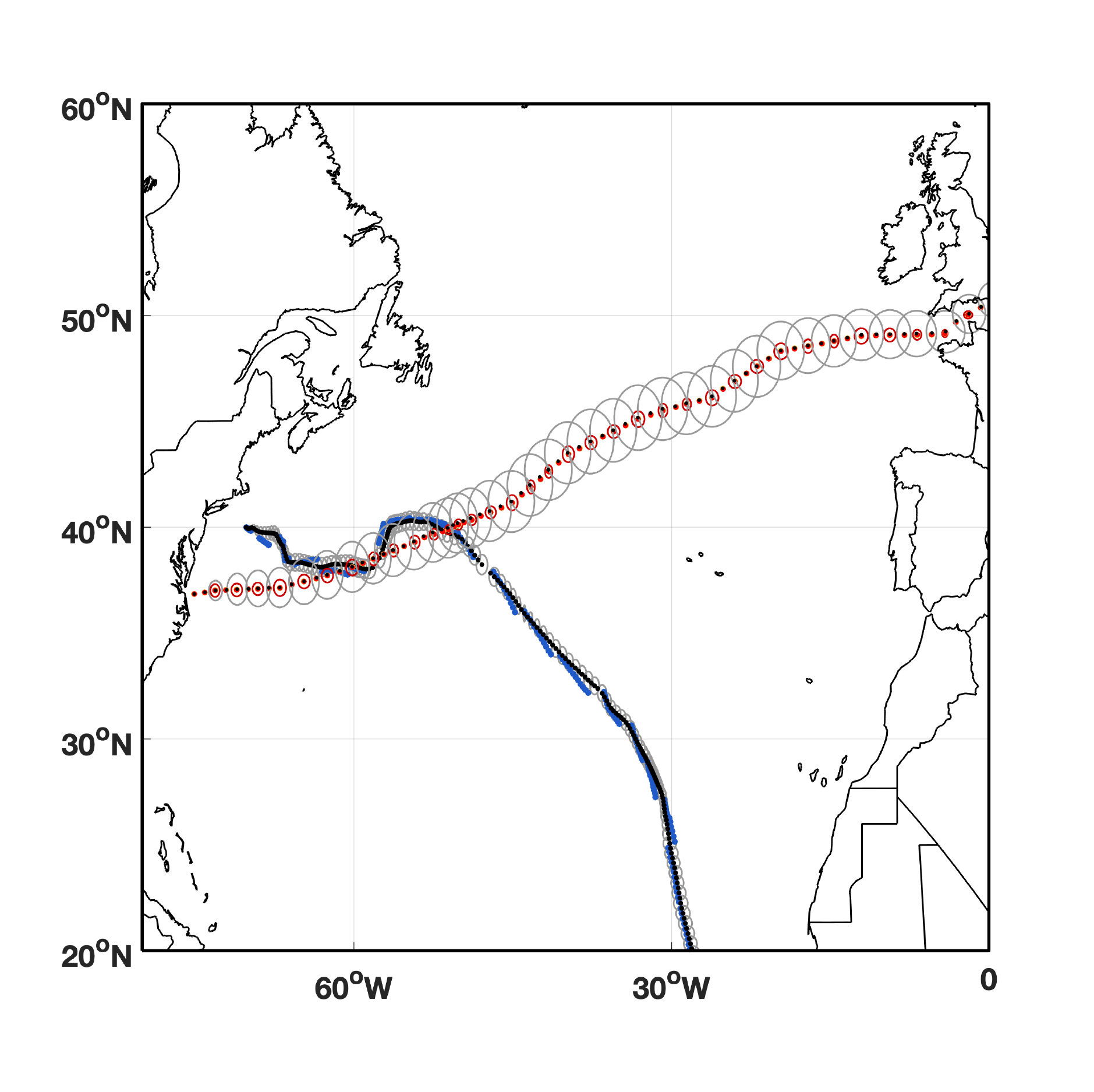}
\caption{The posterior distributions of ship positions. For HQ2 track No.30, blue dots are reported ship positions, black dots represent the posterior means, and circles show posterior uncertainties (one standard deviation). For LQ4 track No.108,  yellow dots are reported ship positions, red dots and circles are the posterior means and uncertainties when a celestial correction happens every midnight ($p = 1$), and black dots and gray circles are the posterior means and uncertainties when there are no celestial corrections ($p = 0$).}
\label{fig:ThreeTypesResults}
\end{figure*}

\begin{figure*}
\centering
\includegraphics[width=25pc]{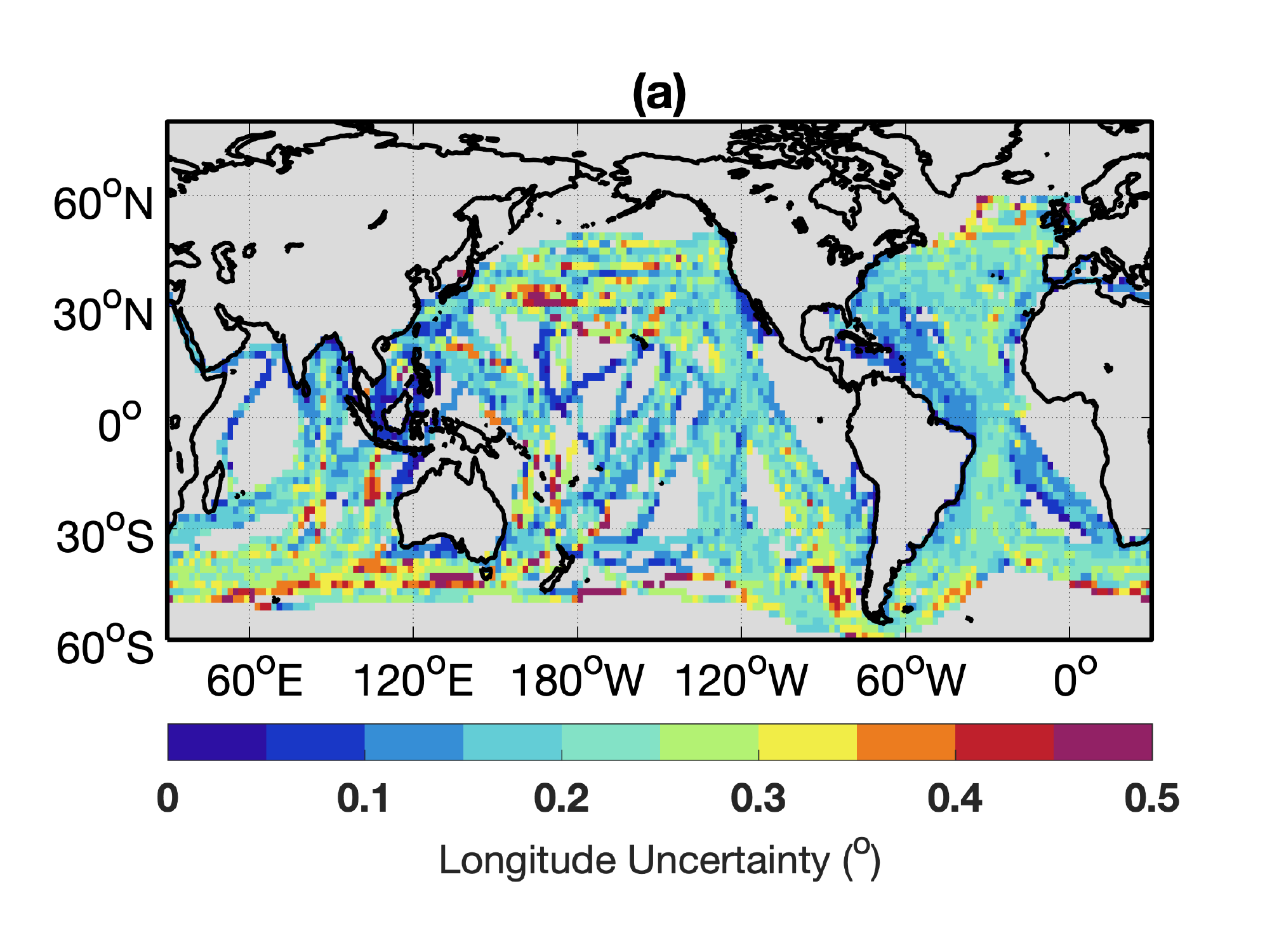}
\includegraphics[width=25pc]{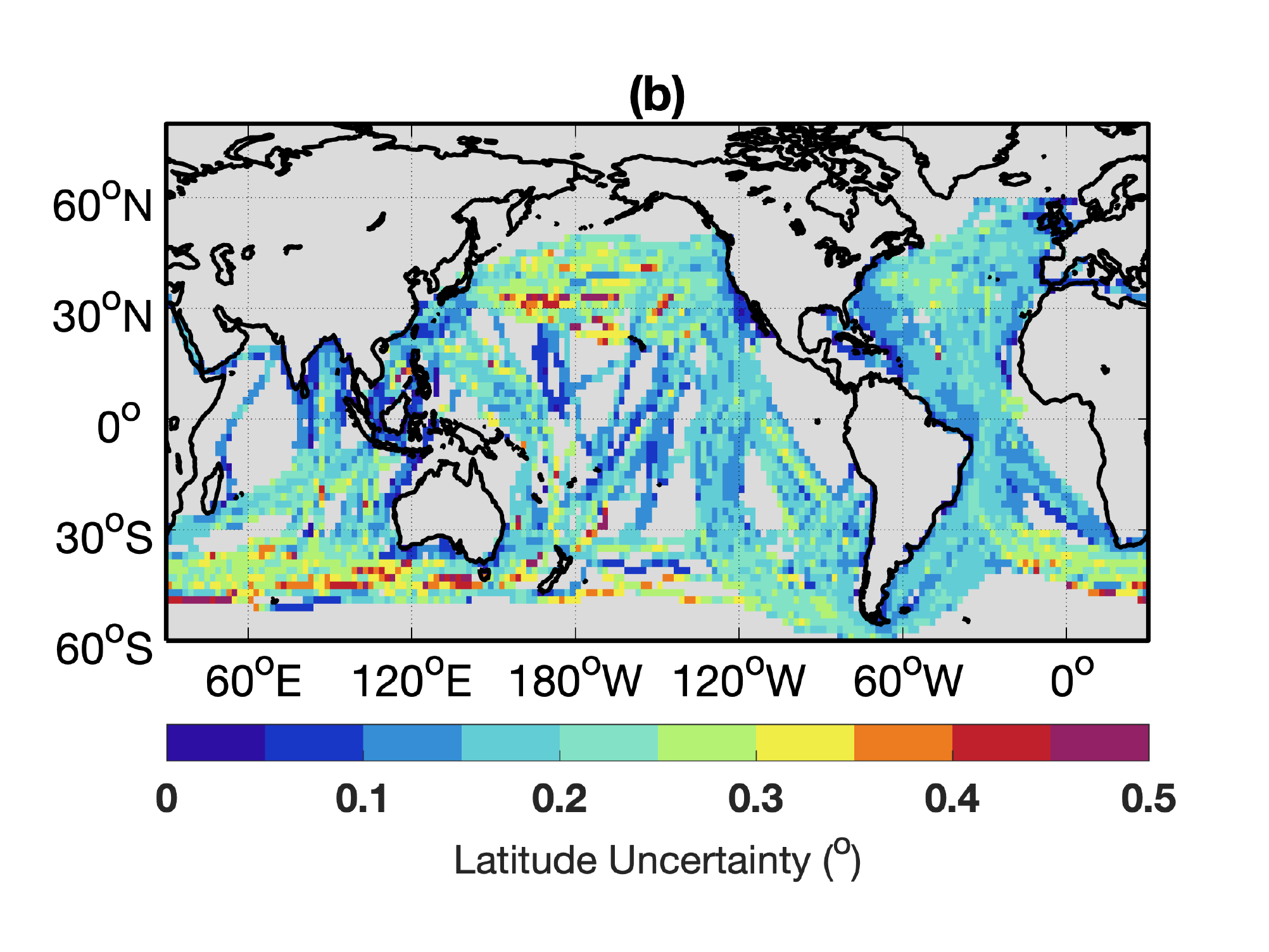}
\caption{Global pattern of random position uncertainties for HQ2 tracks. Individual panels are (a) longitude and (b) latitude.
Track-based estimates of random position uncertainties are gridded to 2$^\circ$ resolution for visualization. Binning is accomplished by averaging random position uncertainties within a grid box in quadrature.}
\label{fig:Good2hourlyPattern}
\end{figure*}

A single track, HQ2 track No.30, is described for purposes of illustrating the results.  HQ2 track No.30 moved from West to East, where blue dots in Figure \ref{fig:ThreeTypesResults} represent the reported ship positions, black dots represent the posterior mean positions, and ellipses indicate posterior one-standard deviation uncertainties.  
The trajectory of the posterior mean tends to diverge from the reported ship positions when celestial navigation updates are large because, unlike reported positions, posterior means take into account not only information from preceding celestial updates but also later ones. Figure \ref{fig:Good2hourlyPattern} demonstrates the global pattern of random position uncertainties for HQ2 tracks.

For celestial correction, the population median of the uncertainty in the longitudinal direction is approximately 33.1 km, or $0.30^\circ$ on the equator, while the population median of the uncertainty in the latitudinal direction is smaller, approximately 24.4 km, or $0.22^\circ$ (see Table \ref{table:BayesHierarResult}). This accords with expectation because celestial correction in the longitudinal direction is subject to errors of both celestial observations and chronometers, whereas celestial correction in the latitudinal direction is free of chronometer errors \citep{bowditch1906american}.  The population median of the uncertainty in the relative ship speed $\hat{s}_t/s_t$ is approximately 19.2\%, which is possibly attributed to the less reliable instruments, e.g., chip logs and patent logs, used in early navigations.
The population median of the uncertainty in ship heading is approximately 0.23 radian, or $13.2^\circ$. 
Among all HQ2 tracks, the 25\%, 50\%, and 75\% quantiles of the absolute difference between the reported ship heading and the posterior mean heading are $2.9^\circ$, $7.9^\circ$ and $20.4^\circ$, respectively. 

\begin{table*}
\begin{minipage}{\textwidth}
\centering
\caption{The population median of navigational parameters.}
\begin{tabular}{|l|c|c|c|c|c|c|c|}
\hline
Quantiles & 5\% & 25\% & 50\% & 75\% & 95\% & std\\
\hline
$\mu_{\tau_x}$\blfootnote{$\mu_{\tau_x}$ and $\mu_{\tau_y}$ denote the population median of the uncertainty in celestial correction along the longitudinal and the latitudinal directions, respectively. $\mu_{\tau_s}$ and $\mu_{\tau_\theta}$ denote the population median of the uncertainty in the relative ship speed $\hat{s}_t/s_t$ and ship heading, respectively.} (km) & 31.6 & 32.4 & 33.1 & 33.7 & 34.6 & 0.92\\
\hline
$\mu_{\tau_y}$ (km) & 23.2 & 23.9 & 24.4 & 24.9 & 25.6 & 0.74\\
\hline
$\mu_{\tau_s}$\hspace{0.1cm} (\%) & 18.3 & 18.8 & 19.2 & 19.6 & 20.1 & 0.5\\
\hline
$\mu_{\tau_\theta}$ \hspace{0.01cm}(rad) & 0.21 & 0.22 & 0.23 & 0.24 & 0.25 & 0.01\\
\hline
\end{tabular}
\label{table:BayesHierarResult}
\end{minipage}
\end{table*}

\begin{table}
\begin{minipage}{\textwidth}
\centering
\caption{Random and systematic position uncertainties.}
\begin{tabular}{ |l|c|c|c|c|}
\hline
\multicolumn{5}{c}{Random position uncertainty} \\
\hline
Quantiles & 25\% & 50\% & 75\% & mean \\
\hline
HQ2 longitude & $0.11^\circ$(12 km) &  $0.18^\circ$(20 km) & $0.23^\circ$(26 km) & $0.18^\circ$(20 km)\\
\hline
HQ2 latitude   & $0.10^\circ$(11 km) &  $0.15^\circ$(17 km) & $0.20^\circ$(22 km) & $0.16^\circ$(18 km)\\
\hline
LQ4 longitude (1.0)\blfootnote{1.0, 0.5 and 0.0 correspond to the the best-case scenario, the random-guess scenario, and the worst-case scenario for LQ4 tracks, respectively.\label{note1}} & $0.11^\circ$(12 km) &  $0.16^\circ$(18 km) & $0.20^\circ$(22 km) & $0.16^\circ$(18 km)\\
\hline
LQ4 longitude (0.5) & $0.16^\circ$(18 km) &  $0.22^\circ$(24 km) & $0.28^\circ$(31 km) & $0.23^\circ$(26 km)\\
\hline
LQ4 longitude (0.0) & $0.35^\circ$(39 km) &  $0.50^\circ$(56 km) & $0.66^\circ$(73 km) & $0.50^\circ$(56 km)\\
\hline
LQ4 latitude\ \ \ (1.0) & $0.09^\circ$(10 km) &  $0.13^\circ$(14 km) & $0.17^\circ$(19 km) & $0.13^\circ$(14 km)\\
\hline
LQ4 latitude\ \ \ (0.5) & $0.13^\circ$(14 km) &  $0.18^\circ$(20 km) & $0.24^\circ$(27 km) & $0.19^\circ$(21 km)\\
\hline
LQ4 latitude\ \ \ (0.0) & $0.29^\circ$(32 km) &  $0.44^\circ$(49 km) & $0.60^\circ$(67 km) & $0.44^\circ$(49 km)\\
\hline
\hline
\multicolumn{5}{c}{Systematic position uncertainty} \\
\hline
Quantiles & 25\% & 50\% & 75\% & mean \\
\hline
HQ2 longitude & $0.06^\circ$(6.7 km) &  $0.16^\circ$(18 km) & $0.34^\circ$(38 km) & $0.24^\circ$(27 km)\\
\hline
HQ2 latitude   & $0.05^\circ$(5.6 km) &  $0.14^\circ$(16 km) & $0.30^\circ$(33 km) & $0.21^\circ$(23 km)\\
\hline
LQ4 longitude (1.0) & $0.00^\circ$(0.0 km) &  $0.00^\circ$(0.0 km) & $0.01^\circ$(1.1 km) & $0.00^\circ$(0.0 km)\\
\hline
LQ4 longitude (0.5) & $0.00^\circ$(0.0 km) &  $0.01^\circ$(1.1 km) & $0.01^\circ$(1.1 km) & $0.01^\circ$(1.1 km)\\
\hline
LQ4 longitude (0.0) & $0.01^\circ$(1.1 km) &  $0.03^\circ$(3.3 km) & $0.06^\circ$(6.7 km) & $0.05^\circ$(5.6 km)\\
\hline
LQ4 latitude\ \ \ (1.0)  & $0.00^\circ$(0.0 km) &  $0.00^\circ$(0.0 km) & $0.01^\circ$(1.1 km) & $0.00^\circ$(0.0 km)\\
\hline
LQ4 latitude\ \ \ (0.5)  & $0.00^\circ$(0.0 km) &  $0.01^\circ$(1.1 km) & $0.01^\circ$(1.1 km) & $0.01^\circ$(1.1 km)\\
\hline
LQ4 latitude\ \ \ (0.0)  & $0.01^\circ$(1.1 km) &  $0.03^\circ$(3.3 km) & $0.06^\circ$(6.7 km) & $0.05^\circ$(5.6 km)\\
\hline
\hline
\multicolumn{5}{c}{Overall position uncertainty} \\
\hline
Quantiles & 25\% & 50\% & 75\% & mean \\
\hline
HQ2 longitude & $0.16^\circ$(18 km) & $0.26^\circ$(29 km) & $0.42^\circ$(47 km) & $0.32^\circ$(36 km)\\
\hline
HQ2 latitude   & $0.14^\circ$(16 km) & $0.23^\circ$(26 km) & $0.37^\circ$(41 km) & $0.28^\circ$(31 km)\\
\hline
LQ4 longitude (1.0) & $0.12^\circ$(13 km) & $0.16^\circ$(18 km) & $0.20^\circ$(22 km) & $0.16^\circ$(18 km)\\
\hline
LQ4 longitude (0.5) & $0.16^\circ$(18 km) & $0.22^\circ$(24 km) & $0.28^\circ$(31 km) & $0.22^\circ$(24 km)\\
\hline
LQ4 longitude (0.0) & $0.36^\circ$(40 km) & $0.51^\circ$(57 km) & $0.68^\circ$(75 km) & $0.51^\circ$(57 km)\\
\hline
LQ4 latitude\ \ \ (1.0) & $0.10^\circ$(11 km) & $0.14^\circ$(16 km) & $0.17^\circ$(19 km) & $0.14^\circ$(16 km)\\
\hline
LQ4 latitude\ \ \ (0.5) & $0.13^\circ$(14 km) & $0.18^\circ$(20 km) & $0.24^\circ$(27 km) & $0.19^\circ$(21 km)\\
\hline
LQ4 latitude\ \ \ (0.0) & $0.30^\circ$(33 km) & $0.46^\circ$(51 km) & $0.62^\circ$(69 km) & $0.46^\circ$(51 km)\\
\hline
\end{tabular}
\label{tab:uncertainty_offset}
\end{minipage}
\end{table}

\subsection{Low-quality four-hourly ship tracks}
As noted, we are unable to infer when LQ4 tracks have their positions updated by celestial observations and, therefore, explore three scenarios wherein celestial positions are made every midnight, with 0.5 probability, or never. Figure \ref{fig:ThreeTypesResults} shows the posterior distributions of LQ4 track No.108 under the best-case and worst-case scenarios.   Under the best-case scenario all the midnight positions along the track are considered being celestially corrected, and LQ4 track exhibits small Brownian-bridge uncertainty structures between consecutive midnight positions.
Under the worst-case scenario we assume no celestial corrections, and LQ4 track exhibits a Brownian-bridge uncertainty structure that spans the departure and arrival points.

Table \ref{tab:uncertainty_offset} summarizes position errors of all LQ4 tracks under the three scenarios. The overall position uncertainty (MSE) combines the random position uncertainty (standard deviation) and the systematic position uncertainty (bias) using the bias-variance decomposition. On average, random position uncertainties of LQ4 tracks are $0.50^\circ$ (56 km on the equator) in longitude and $0.44^\circ$(49 km) in latitude under the worst-case scenario, and $0.16^\circ$ (18 km) in longitude and $0.13^\circ$(14 km) in latitude under the best-case scenario. The half-probability scenario is similar to, albeit of course slightly more uncertain, than the best-case scenario. 
The random position uncertainty is represented as better than HQ2 tracks, where the latter have only 87\% of nights associated with a celestial correction.  
Because there are no apparent jumps, LQ4 tracks are inferred to have smaller systematic uncertainties than estimated for HQ2 tracks.

\begin{figure*}
\centering
\includegraphics[width=14pc]{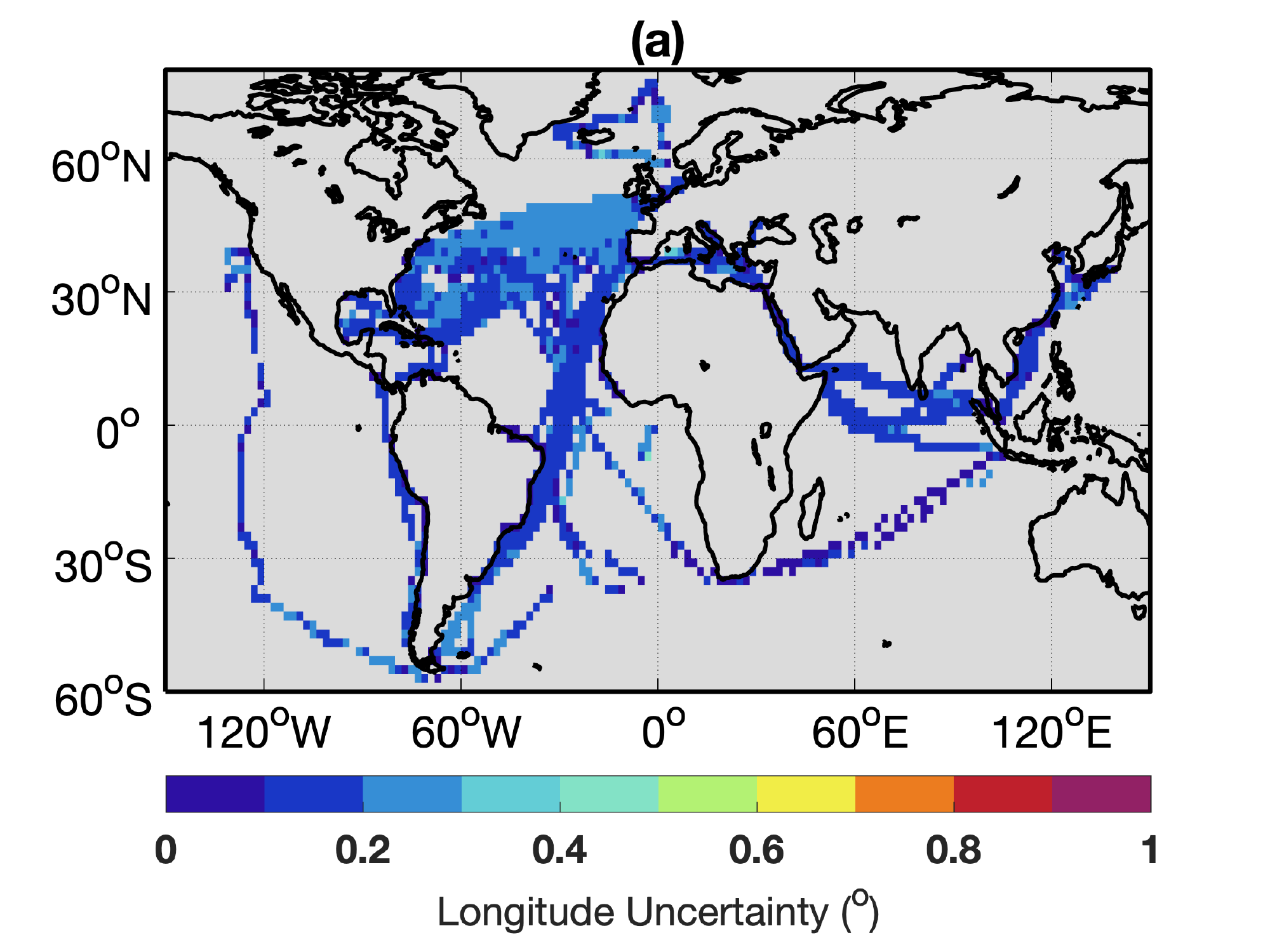}
\includegraphics[width=14pc]{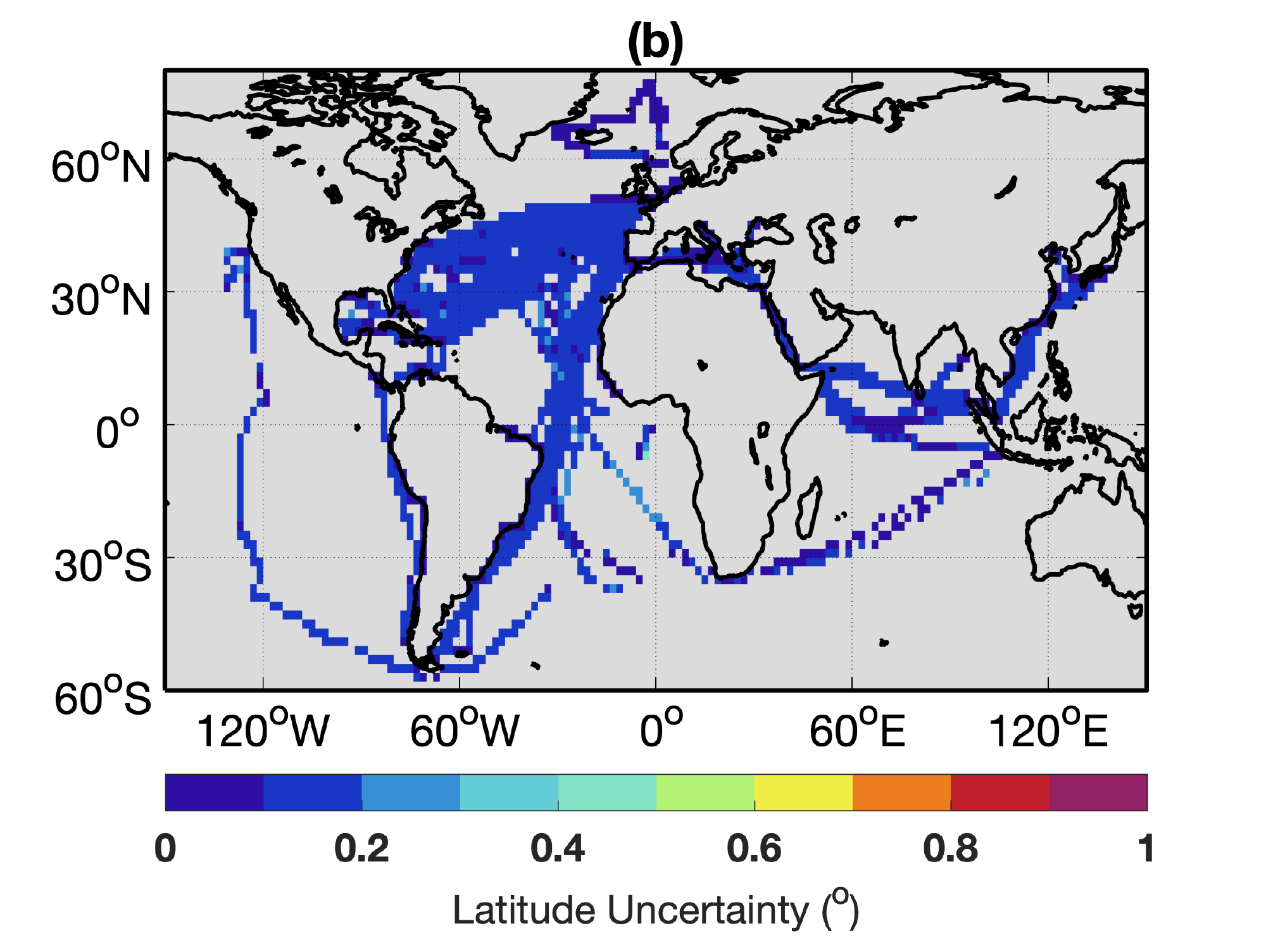}
\includegraphics[width=14pc]{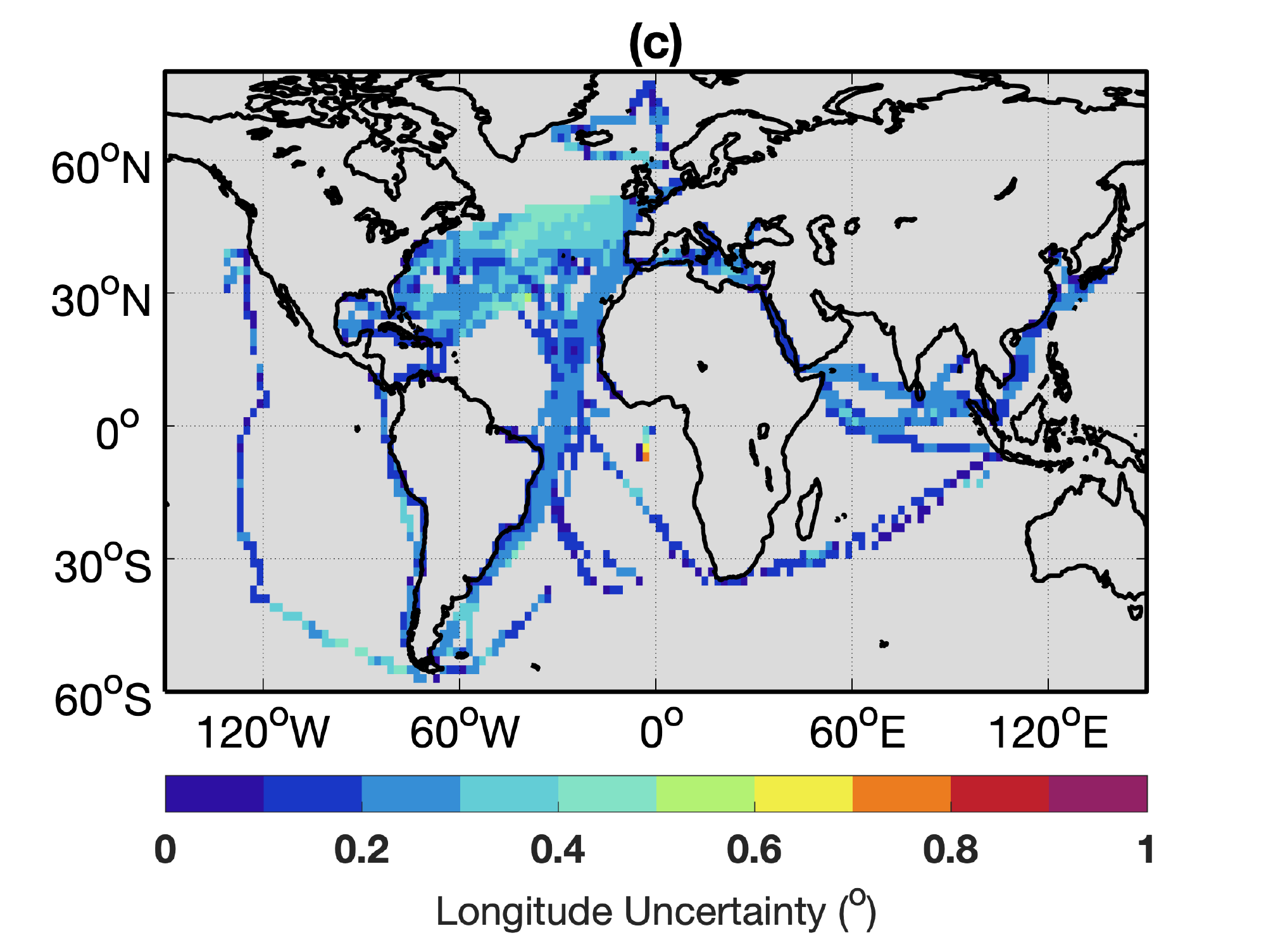}
\includegraphics[width=14pc]{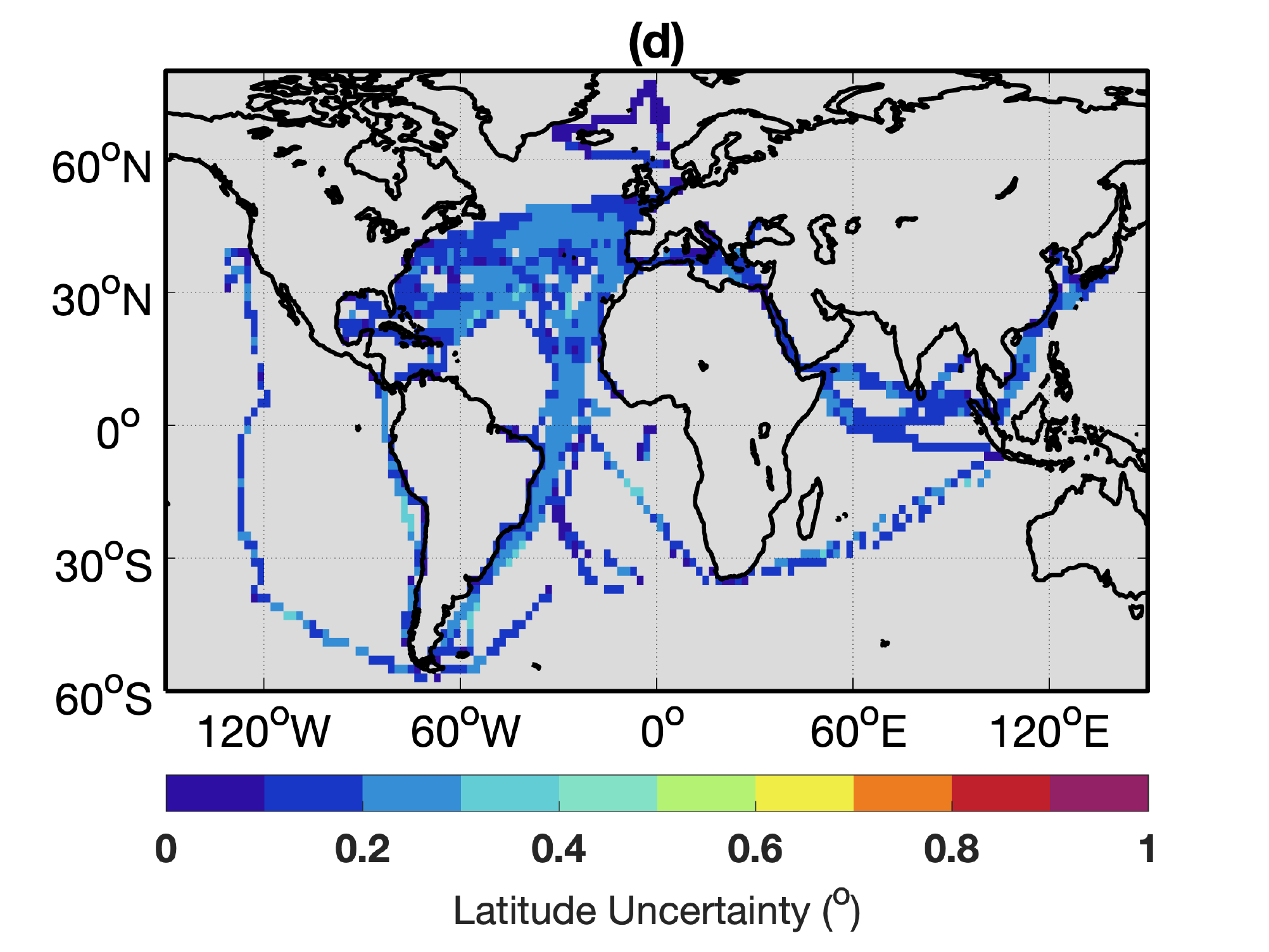}
\includegraphics[width=14pc]{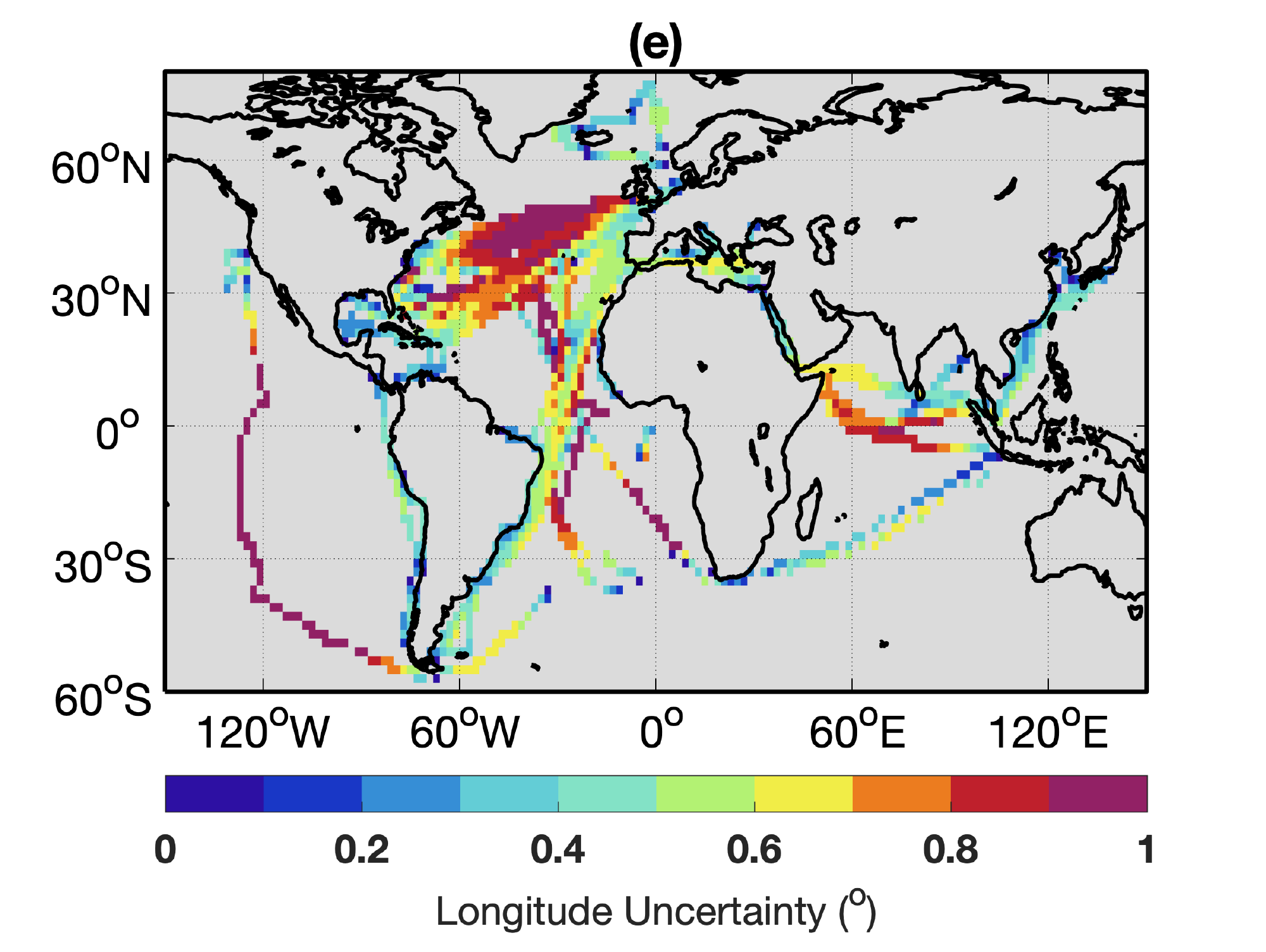}
\includegraphics[width=14pc]{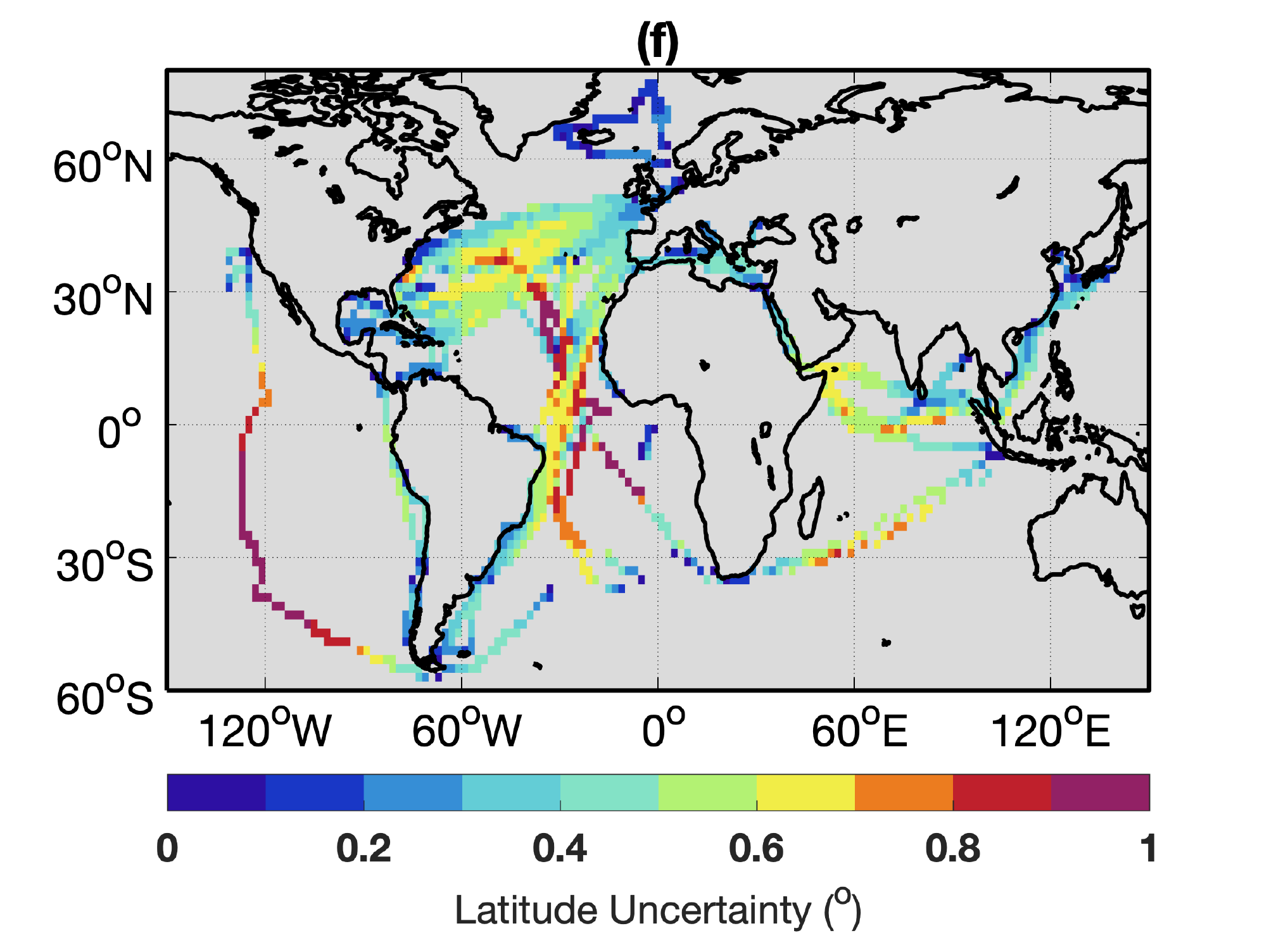}
\caption{Global pattern of random position uncertainties for LQ4 tracks.  Left panels show longitude uncertainties for (a) best-case, (b) random-guess, and (c) worst-case scenarios. Right panels are as the left but for latitude uncertainties.  In the best-case scenario, a celestial correction happens at each midnight.  In the random-guess scenario, midnight positions have a probability of 0.5 to be corrected, whereas in the worst-case scenario, celestial corrections never happen.  Maps are shifted to center on the Atlantic for the purpose of visualization.  Procedures of generating maps from individual measurements are as per Figure \ref{fig:Good2hourlyPattern}.}
\label{fig:OthersPattern}
\end{figure*}

The Brownian-bridge uncertainty structure implies larger errors associated with longer journeys and being further away from departure and arrival points, such that positions in the interior of oceans are generally more uncertain (see Figure \ref{fig:OthersPattern}).  
Note that position uncertainties depend not only on the distance from coasts or islands, but also on directions that ships are heading, which determines the relative magnitude of the uncertainties in longitude and latitude.

%%%%%%%%%%%%%%%%%%%%%%%%%%%%%%%%%%%%%%%%%%%%%%%%%%%%%%%%%%%%%%%%%%%%%%%%%%%%%
%%%%%%%%%%%%%%%%%%%%%%%%%%%%%%%%%%%%%%%%%%%%%%%%%%%%%%%%%%%%%%%%%%%%%%%%%%%%%
\subsection{SST uncertainties}
\label{subsec:sst_results}

To quantify uncertainties in SSTs associated with errors in position, we sample a high-resolution SST dataset with position errors that mimic those expected from our analysis of HQ2 tracks.  We use the Multi-Scale Ultra-high resolution Sea Surface Temperature dataset (MURSST) \citep{chin2017multi} that incorporates infrared and microwave satellite retrievals and observations from ships and buoys.  Although the data is obviously more recent than the 1885 ship tracks that we analyze, MURSST has the advantage of having a 0.01$^\circ$ spatial resolution that is comparable to the HQ2 ship-track precision.  Estimated SST uncertainties are still meaningful because the basic SST patterns --- including those related to equator-to-pole temperature gradients, boundary currents, gyres, and upwelling regions --- are stable features of the ocean circulation \citep{wunsch2004gulf}.

SSTs in MURSST are repeatedly sampled in order to estimate uncertainties.  
For each posterior ship track realized, we sample SSTs at the realized positions in MURSST on the corresponding high-resolution monthly climatology over 2003-2018. In total we obtain 1,000 posterior ship tracks and their corresponding SSTs are sampled. The uncertainties in SSTs are estimated by taking the standard deviation across these 1,000 samples.
For purposes of visual display, the uncertainties are re-gridded at 2$^\circ$ resolution.  An assumption associated with the usage of daily SST products is that diurnal SST signals are relatively uniform at small spatial scales.  This assumption is, in general, valid in cold seasons, when diurnal cycles are small \citep{morak2016climatological}, and over the extra-tropical ocean, where stratiform clouds associated with large-scale weather systems dominate.  We note that in the places where diurnal cycles are affected by small-scale convective clouds, such as in parts of the tropics, sampling from daily-average SST products leads to a larger underestimation of SST uncertainties.

\begin{figure*}
\centering
\includegraphics[width=25pc]{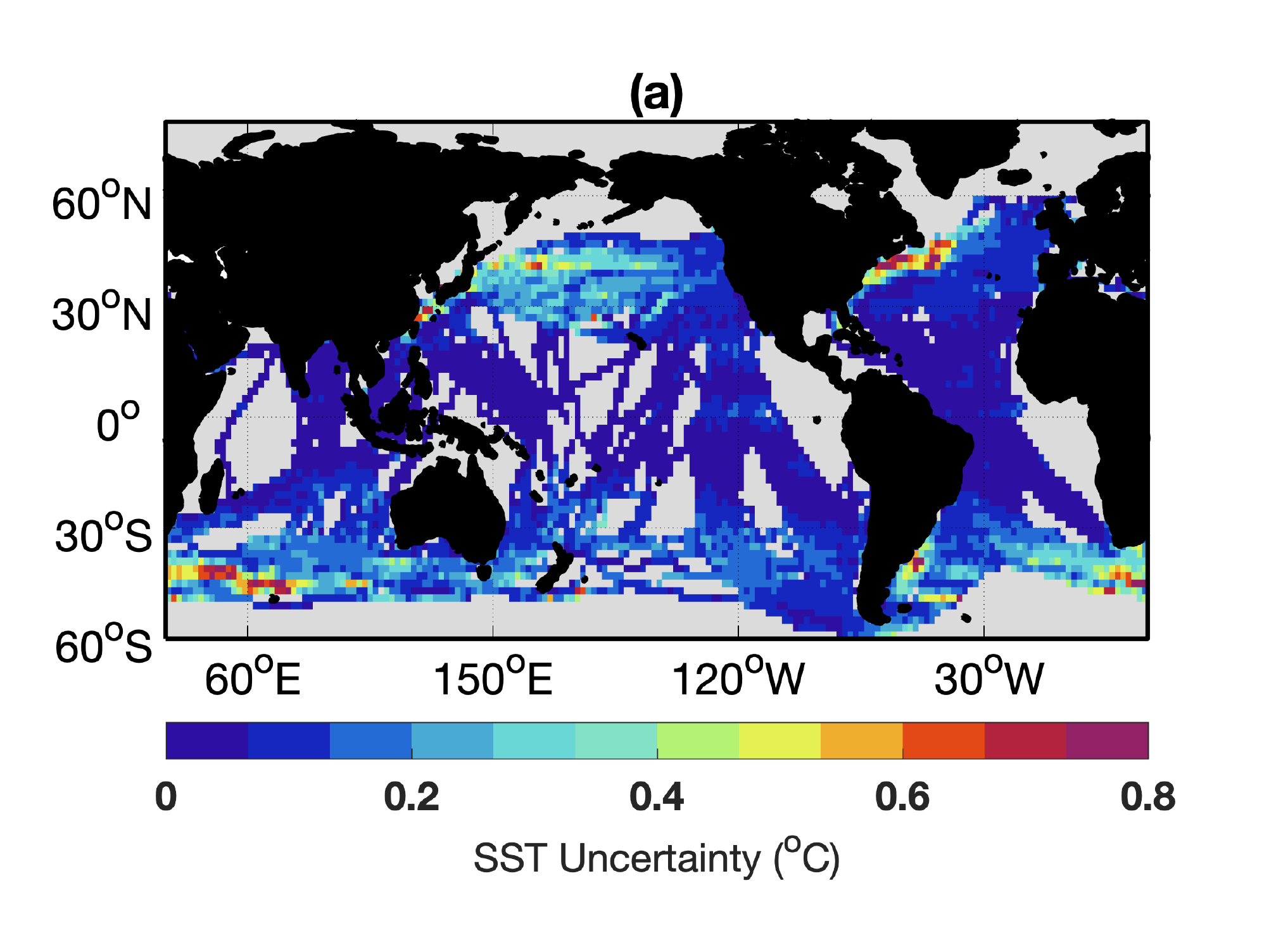}
\includegraphics[width=25pc]{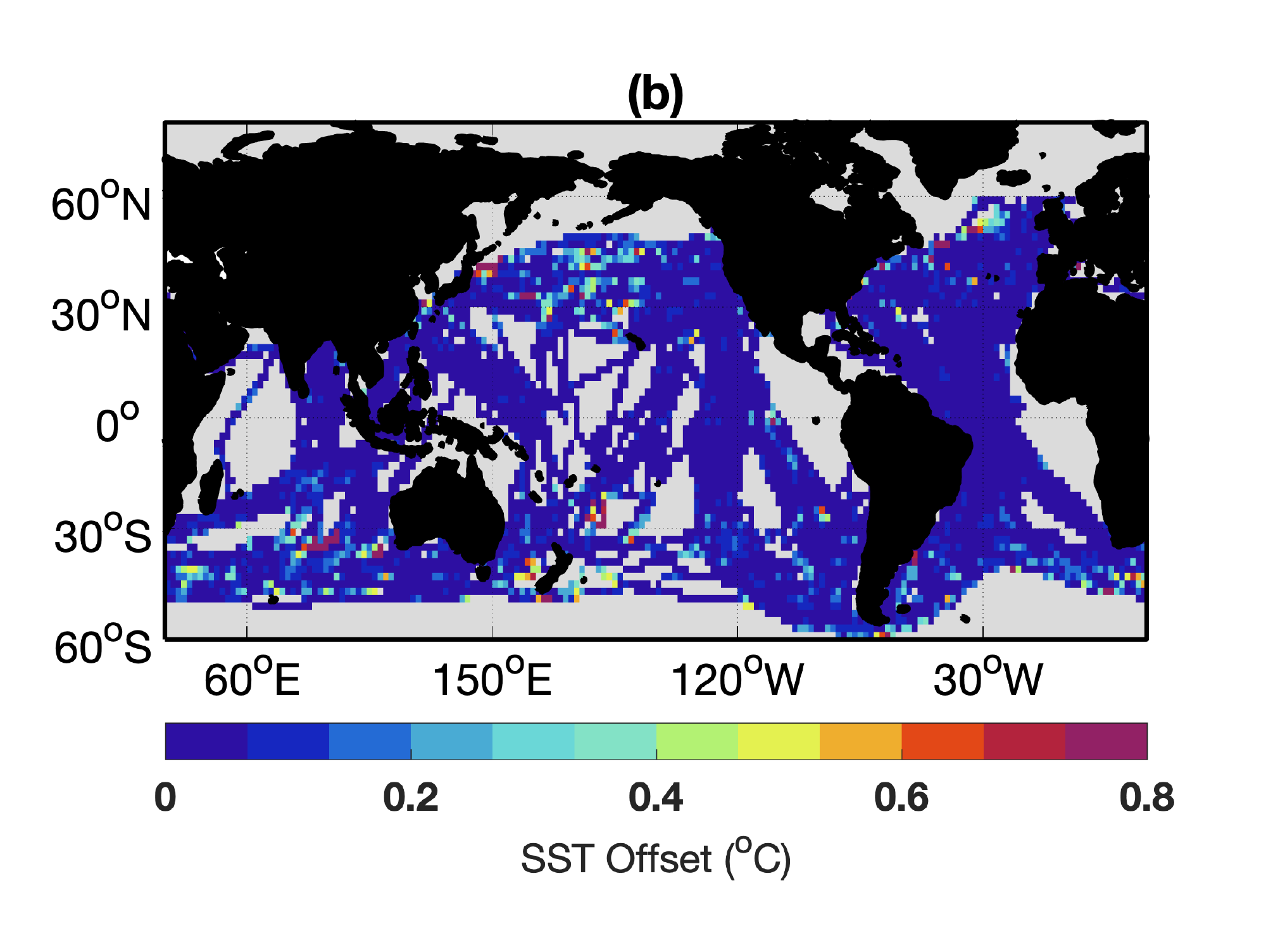}
\caption{SST uncertainties associated with position errors for HQ2 tracks.  Individual panels are (a) random SST uncertainties and (b) systematic SST offsets.  Results are binned to $2^\circ$ grids for visualization. }
\label{fig:SSTJan}
\end{figure*}

\begin{figure*}
\centering
\includegraphics[width=14pc]{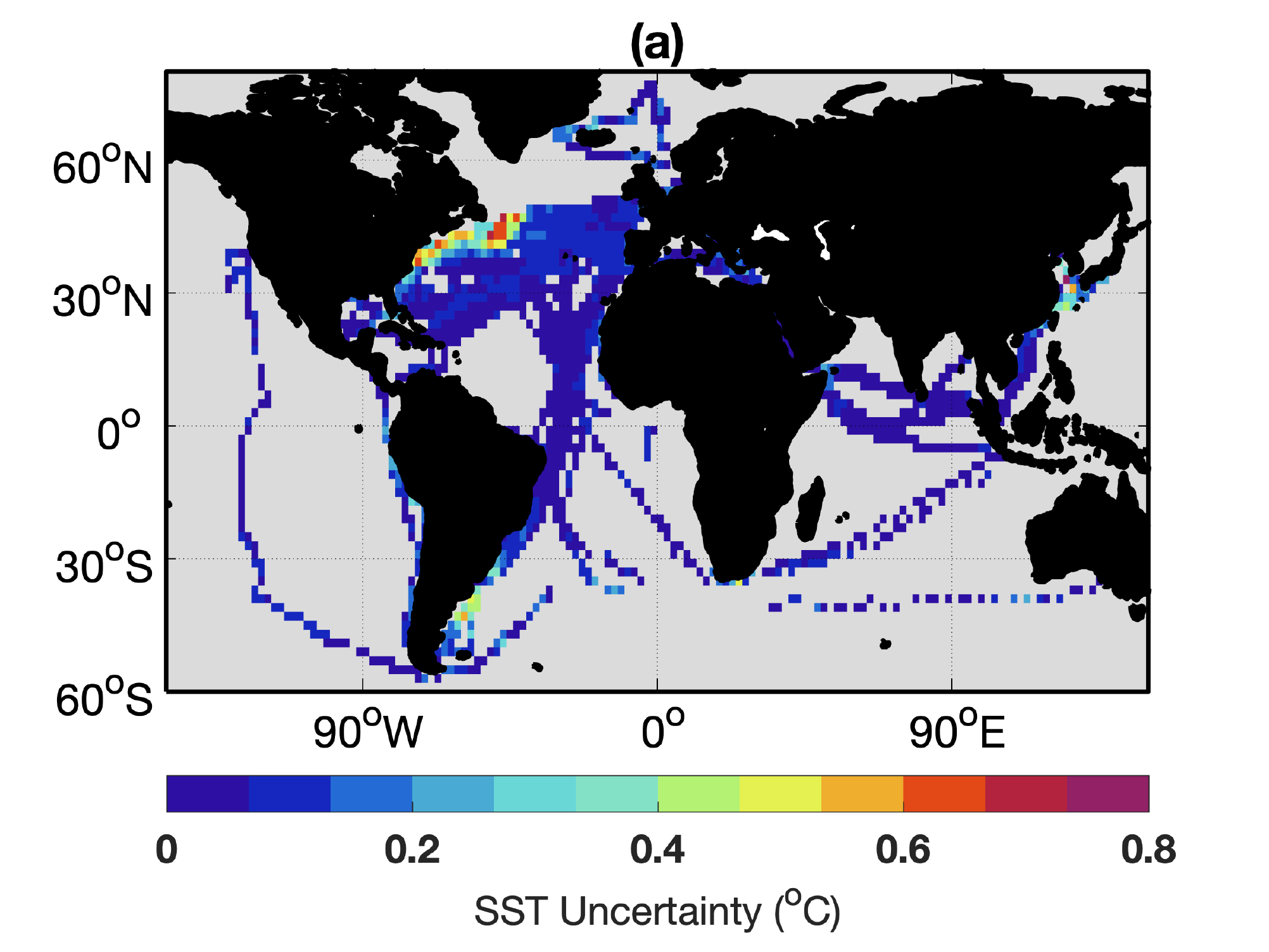}
\includegraphics[width=14pc]{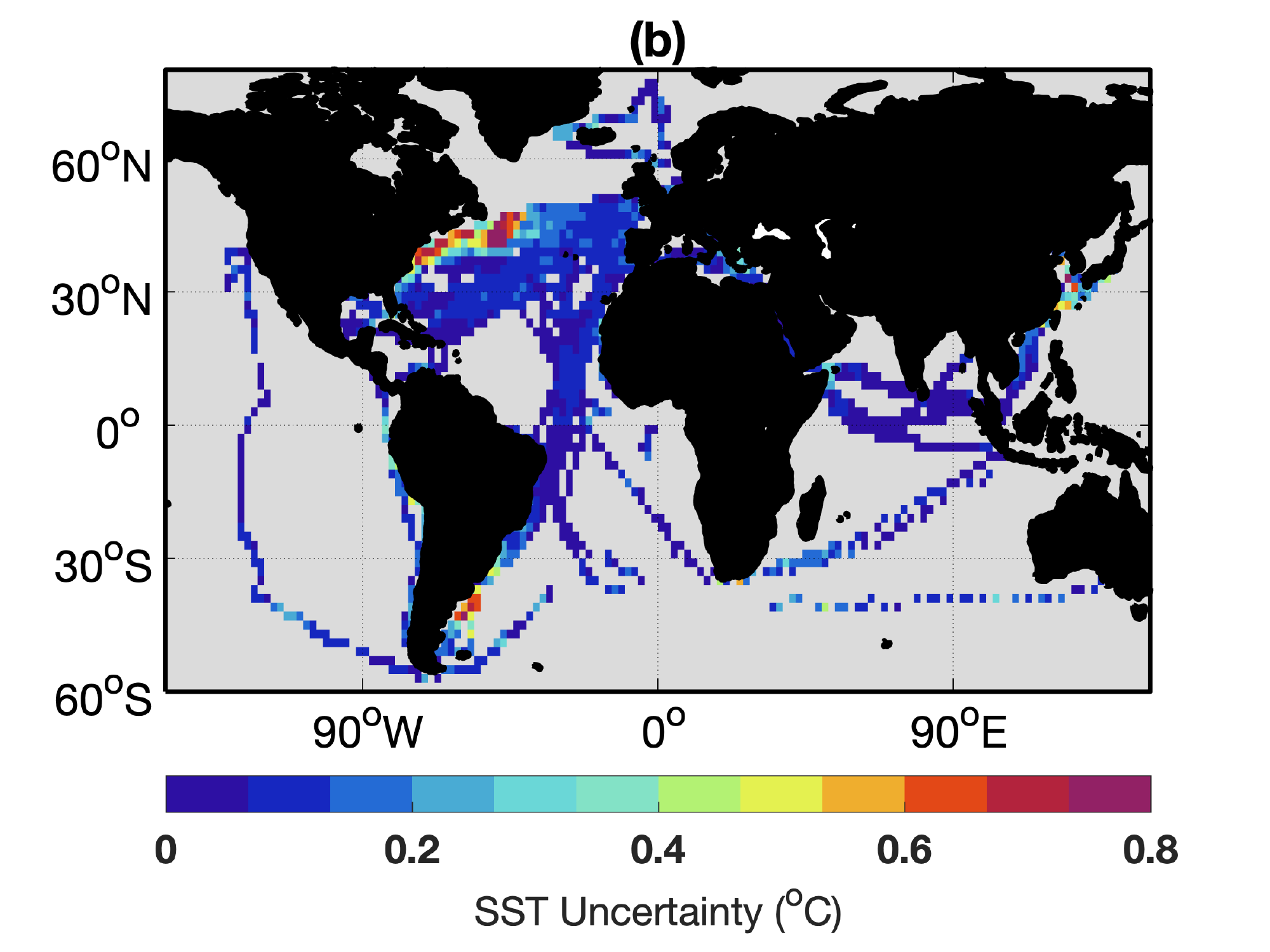}
\includegraphics[width=14pc]{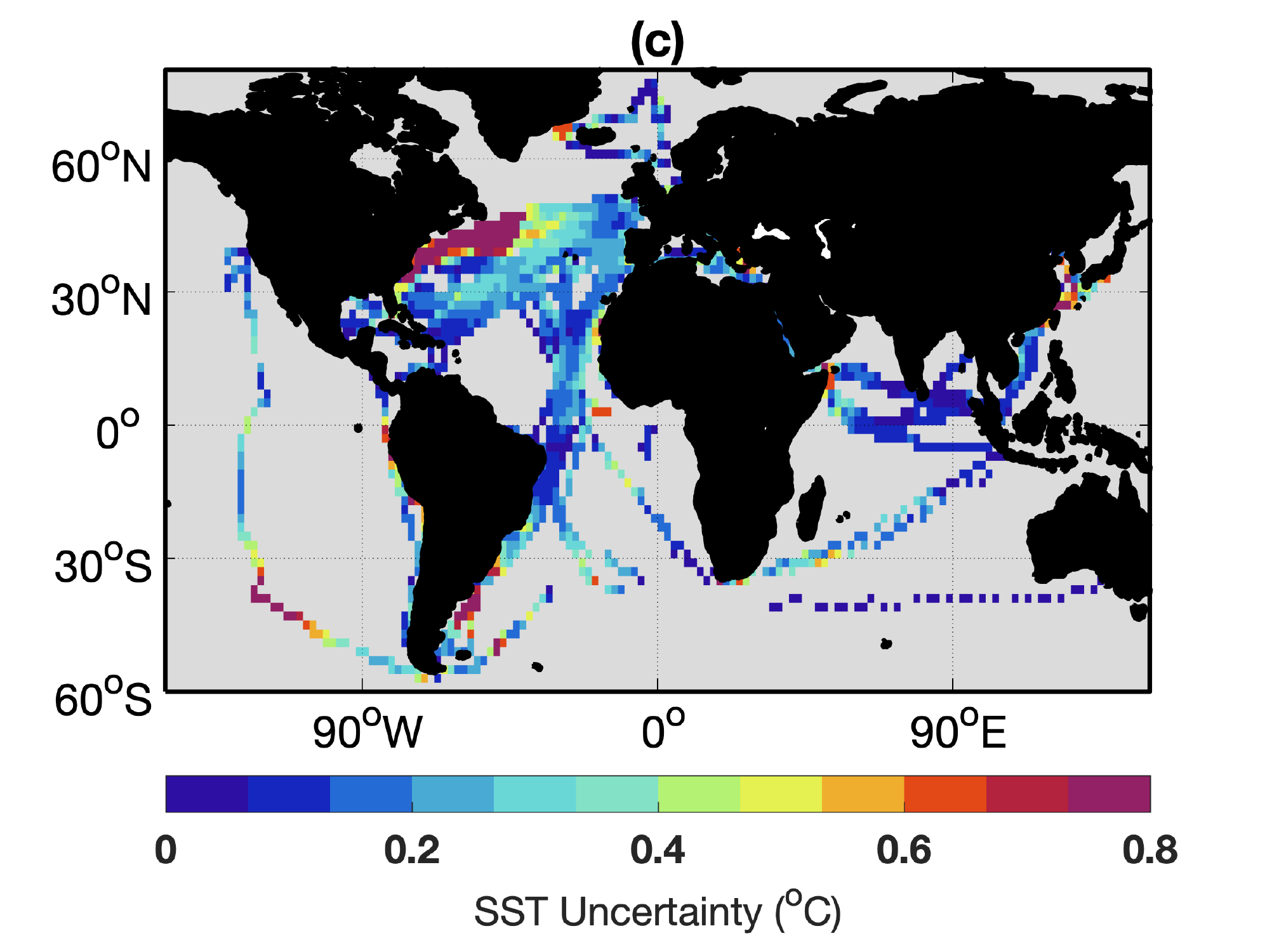}
\includegraphics[width=14pc]{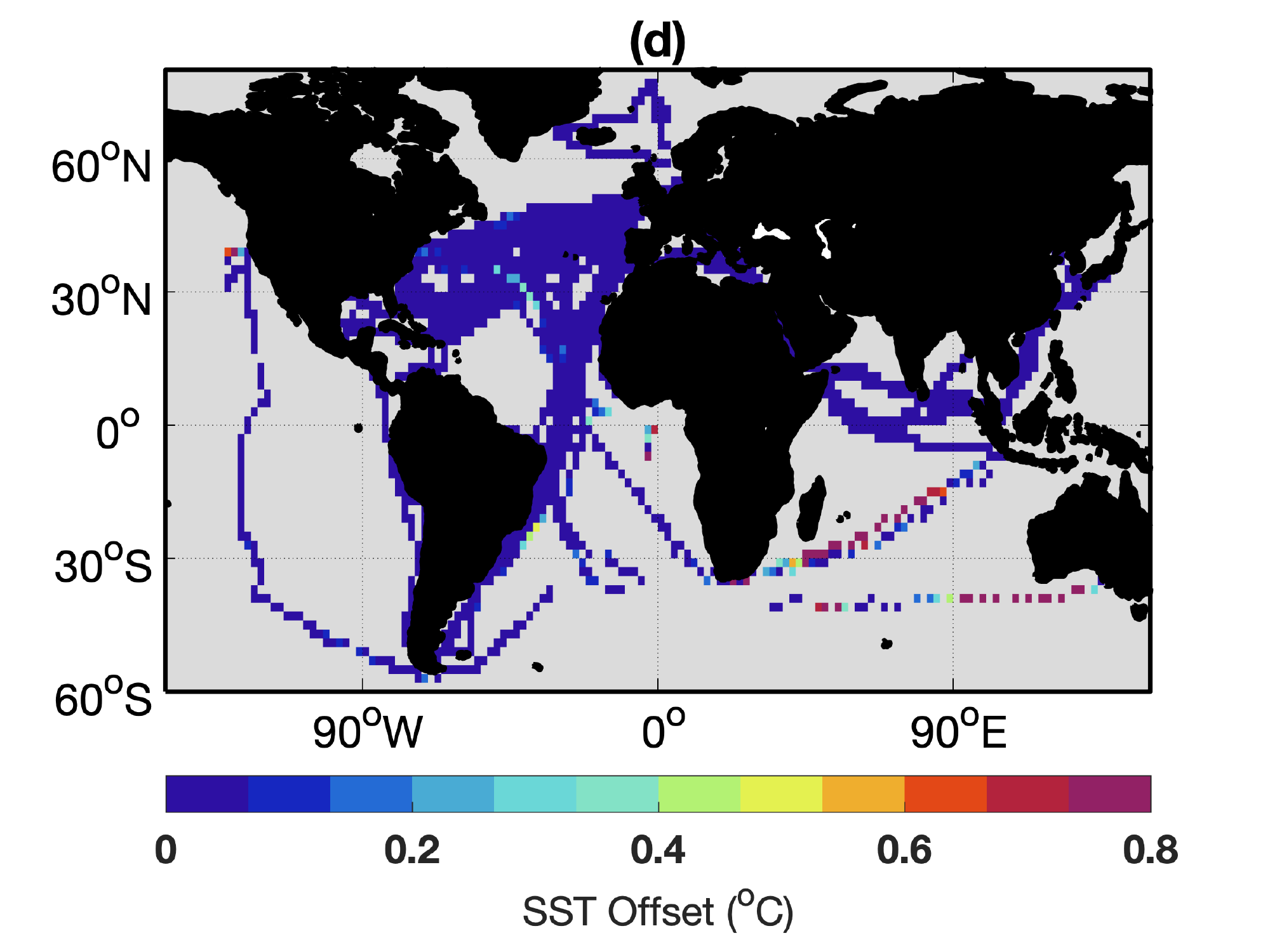}
\caption{SST uncertainties associated with position errors for LQ4 tracks.  Panels (a), (b), and (c) demonstrate random SST uncertainties under the best-case scenario, the random-guess scenario, and the worst-case scenario, respectively.  Panel (d) shows SST offsets under the best-case scenario.  Results are binned to $2^\circ$ grids for visualization.}
\label{fig:SSTJan4hr}
\end{figure*}

Position errors induce uncertainties in SST in regions where position errors are large and SST gradients are strong.  On average, position errors in HQ2 tracks translate into $0.11^\circ$C SST uncertainties (see Table \ref{tab:SST_uncertainty_offset}), but can be as large as 0.33$^\circ$C in the Northwest Atlantic where the western boundary current detaches from the East Coast of the US (see Figure \ref{fig:SSTJan}).  Larger SST uncertainties are also found in regions where other boundaries currents detach in the Northwest Pacific and Southwest Atlantic, as well as in the vicinity of the Aghulas Current south of South Africa.  For LQ4 tracks (see Table  \ref{tab:SST_uncertainty_offset} and Figure \ref{fig:SSTJan4hr}), SST uncertainties average $0.22^\circ$C and reach $0.94^\circ$C in the mid-latitude Northwest Atlantic under the zero-celestial observation scenario, and are similar to HQ2 tracks in the other scenarios.
\begin{table}
\begin{minipage}{\textwidth}
\caption{A comparison of SST uncertainty and SST offset.}
\centering
\begin{tabular}{ |l|c|c|c|c||c|c|c|c|}
\hline
Global\footnote{The summary statistics are calculated using SSTs at all positions along ship tracks.}  & \multicolumn{4}{|c||}{SST uncertainty ($^\circ$C)} &  \multicolumn{4}{|c|}{SST offset ($^\circ$C)} \\
\hline
Quantiles & 25\% & 50\% & 75\% & mean &  25\% & 50\% & 75\% & mean\\
\hline
HQ2 & 0.03 & 0.06 & 0.12 & 0.11 & -0.05 & 0.01 & 0.07 & 0.02\\
\hline
LQ4 (1.0) & 0.02 & 0.05 & 0.09 & 0.08 & -0.01 & 0.00 & 0.01 & 0.00 \\
\hline
LQ4 (0.5) & 0.03 & 0.07 & 0.12 & 0.10 & -0.01 & 0.00 & 0.01 & 0.00\\
\hline
LQ4 (0.0) & 0.08 & 0.16 & 0.27 & 0.22 & -0.03 & 0.00 & 0.03 & 0.01\\
\hline

Regional\footnote{The summary statistics are calculated using SSTs restricted in the regions over the Gulf stream (280-320$^\circ$E, 40-50$^\circ$N and 280-300$^\circ$E, 35-40$^\circ$N).}  & \multicolumn{4}{|c||}{SST uncertainty ($^\circ$C)} &  \multicolumn{4}{|c|}{SST offset ($^\circ$C)} \\
\hline
Quantiles & 25\% & 50\% & 75\% & mean &  25\% & 50\% & 75\% & mean\\
\hline
HQ2 & 0.08 & 0.23 & 0.46 & 0.33 & -0.22 & -0.00 & 0.24 & 0.03 \\
\hline
LQ4 (1.0) & 0.11 & 0.26 & 0.47 & 0.32 & -0.05 & 0.00 & 0.04 & -0.01\\
\hline
LQ4 (0.5) & 0.17 & 0.37 & 0.62 & 0.43 & -0.07 & -0.01 & 0.04 & -0.01\\
\hline
LQ4 (0.0)  & 0.42 & 0.83 & 1.29 & 0.94 & -0.30 & -0.03 & 0.25 & -0.02\\
\hline
\end{tabular}
\label{tab:SST_uncertainty_offset}
\end{minipage}
\end{table}

%%%%%%%%%%%%%%%%%%%%%%%%%%%%%%%%%%%%%%%%%%%%%%%%%%%%%%%%%%%%%%%%%%%%%%%%%%%%%%%%%%%%%%%%%%%%%%%%
\section{Concluding Remarks}{\label{sec:future}}
We are unaware of previous efforts to quantify the position uncertainty of historical observations from ships. The magnitude of uncertainties of a given quantity associated with position uncertainties will depend on the local gradient of that quantity, which we estimated for SSTs is on the order of $0.1^\circ$C over the globe but can be up to $0.3^\circ$C regionally (Figure \ref{fig:SSTJan}, \ref{fig:SSTJan4hr}, and Table \ref{tab:SST_uncertainty_offset}).  Expected SST errors from position uncertainties are small as compared with other error sources. For example, individual SST measurements have random measurement errors of approximately $1.0^\circ$C and are subject to biases that average approximately $0.4^\circ$C cool when measured using buckets and approximately $0.1^\circ$C warm when collected using the engine-room-intake method \citep{kennedy2011reassessing, kennedy2011reassessingB}. Recent studies also identified significant offsets up to $0.5^\circ$C associated with national groups by comparing collocated measurements \citep{chan2019systematic}.

We note, however, that position uncertainties are distinct from other errors in that they will influence all observations made from the same journey, leading to correlated errors along ship tracks. Moreover, position uncertainties will affect the estimation of biases and random SST errors since many existing approaches require pairing collocated measurements \citep{kent2006toward, chan2019systematic}. As noted, position errors can also lead to changes in tail behavior an high-order moments in grid-box averaged SSTs \citep{director2015connecting}.  We suspect that similar concerns arise with respect to other mapping procedures that interpolate using weighted averages of observations depending upon uncertain positions.

The HQ2 data provides sufficiently frequent and precise observations to characterize uncertainties, but may not be indicative of the overall accuracy of SST positions in 1885.  We speculate the vessels enrolled in the U.S. meteorological program represent a subset of ships wherein a higher priority was placed upon navigation.  Data that is reported with lower resolution and without distinct indications of celestial navigational updates may reflect cruises wherein navigation was a lower priority, or less feasible given limitations with regard to expertise, equipment, or labor.  Thus, estimated position errors may not reflect the overall uncertainty of position data in 1885.  There will also be heterogeneity in reports amongst ships that we have not fully accounted for. Some cruises presumably had lower need of precise navigation; for example, a zonal cross-Atlantic cruise would have less need of determining longitude for purposes of ensuring landfall than a cruise with a meridional heading whose intended port was an island.  In addition, there are also possibilities that longitude and latitude are not celestially corrected simultaneously for earlier navigators because of less-widely deployed ship-board chronometers \citep{bowditch1906american}.

We focus on a single year in developing and testing our procedure, but it would be useful to extend the analysis over a longer time horizon and to a greater fraction of the data.  
In 1885, around 85\% of observations are associated with ship tracks.  Furthermore, \citet{carella2017probabilistic} have provided estimates of additional data belonging to individual ship tracks, bringing the percentage of observations associated with ship tracks in 1885 to 90\%.
In more data-rich intervals, however, distinguishing individual ship track becomes more difficult, such that between 1900-1940 only 60\% of observations, on average, are associated with tracks.

There is presumably a trend toward increasing accuracy of position with time, given technological improvements in marine navigation. This implies that errors and modification of SST distributions introduced through positional error will decrease through time, possibly having consequence for trend estimates, especially those in the vicinity of sharp SST gradients.  Position error may limit the spatial resolution over which trends can accurately be determined.  It would be useful to estimate uncertainties for a gridded SST product with global coverage that, in addition to accounting for observational SST errors \citep{kennedy2011reassessingB} and correcting for biases \citep{kent2017call, chan2019correcting}, also accounted for position errors.

%%%%%%%%%%%%%%%%%%%%%%%%%%%%%%%%%%%%%%%%%%%%%%%%%%%%%%%%%%%%%%%%%%%%%%%%%%%%%
\section*{Acknowledgements}
Natesh Pillai and Chenguang Dai were supported by the Office of Naval Research.
Peter Huybers and Duo Chan were supported by the Harvard Global Institute. 
%%%%%%%%%%%%%%%%%%%%%%%%%%%%%%%%%%%%%%%%%%%%%%%%%%%%%%%%%%%%%%%%%%%%%%%%%%%%%

%%%%%%%%%%%%%%%%%%%%%%%%%%%%%%%%%%%%%%%%%%%%%%%%%%%%%%%%%%%%%%%%%%%%%%%%%%%%%
\section{Appendix}{\label{sec:check}}
To evaluate the fitness of the proposed models, we check the following aspects, including the posterior predictive distributions of HQ2 tracks and the posterior distributions of the navigational parameters. 

\subsection{Posterior predictive check of HQ2 tracks}
The posterior predictive check is a self-consistency check in the sense that any replicated data simulated from the posterior predictive distribution should look similar to the observed data. 

To generate posterior predictive samples of each HQ2 track, we first randomly sample a set of navigational parameters $\tau_x, \tau_y, \tau_s, \tau_\theta$ from their posterior samples. Recall that $\mathcal{C} = \{t_1, \cdots, t_m\}$ is the set of the time steps when celestial updates happen. 
For $k\in[1:(m - 1)]$, if $t = t_k$, that is, the reported ship position contains only celestial observational errors, we randomly sample $p_t^x, p_t^y$ from their posterior samples, and sample $q_t^x, q_t^y$ from
\begin{equation}
q_t^x  \sim N(p_t^x, \left({\tau_x}\cos\psi_t\right)^2),\ \ \ q_t^y  \sim N(p_t^y, {\tau_y}^2).
\end{equation}
Otherwise if $t \in (t_k, t_{k + 1})$, we follow the dead reckoning navigation. That is, we randomly sample $s_t, \theta_t, \beta_k$ from their posterior samples, and sample $\hat{s}_t, \hat{\theta}_t$ from
\begin{equation}
\hat{s}_t  \sim N(s_t, (\tau_ss_t)^2),\ \ \ \ \ \hat{\theta}_t   \sim N(\theta_t + \beta_k, \tau_\theta^2).
\end{equation}
Then we generate $q_t^x, q_t^y$ as follows,
\begin{equation}
\begin{aligned}
q_t^x  = q_{t - 1}^x + 2\hat{s}_t \cos(\hat{\theta}_t), \ \ \ \
q_t^y = q_{t - 1}^y + 2\hat{s}_t \sin(\hat{\theta}_t).
\end{aligned}
\end{equation}

\begin{figure*}
\centering
\includegraphics[width=30pc]{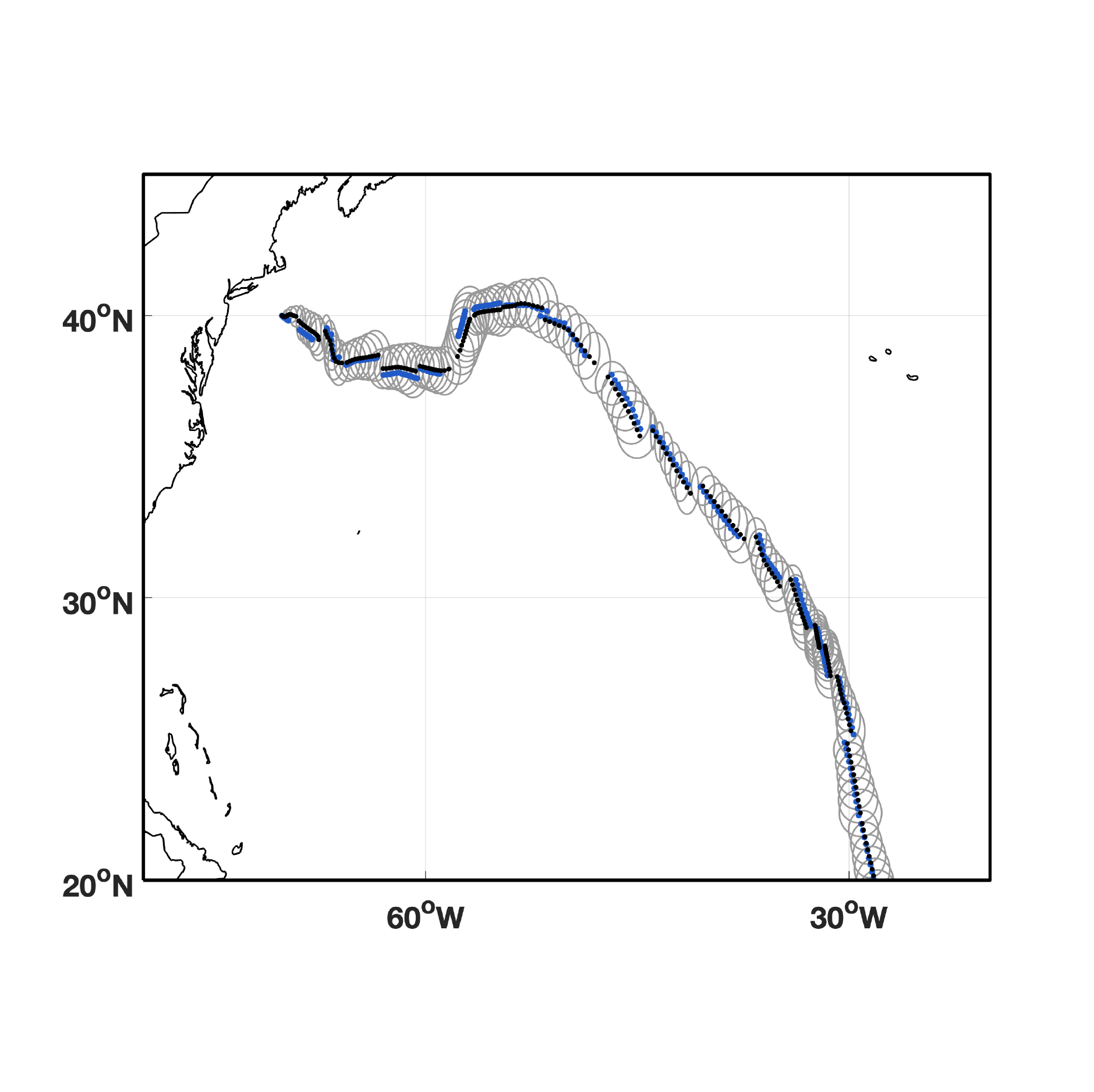}
\caption{Posterior predictive distributions of HQ2 track No.30. The blue dots represent the reported ship positions, the black dots represent the posterior predictive means, and the circles calibrate the posterior predictive uncertainties (one standard deviation).}
\label{fig:ppc}
\end{figure*}

Figure \ref{fig:ppc} shows the posterior predictive distributions of HQ2 track No.30, generated based on 1,000 posterior predictive samples. We see that the empirical means of the posterior predictive samples imitate the observed data. As expected, the position uncertainty structure follows a quasi-daily pattern.

\subsection{Check of navigational uncertainties via a linearized model}

The navigational parameter estimates play a crucial role in the downstream analysis.
We propose an independent linearized model to directly estimate these navigational parameters using HQ2 tracks, and compare the obtained estimates to the results in Table \ref{table:BayesHierarResult}.

We first linearize the dead reckoning process by Taylor expansion and omit the second order terms. For $t \notin \mathcal{C}$, we have
\begin{equation}
\begin{aligned}
\hat{\lambda}^x_t & = \hat{\lambda}^x_{t - 1} + 2s_t  (1 + e_t^s) \cos(\theta_t + e_t^\theta) \\
& \approx \hat{\lambda}^x_{t - 1} + 2s_t\cos\theta_t + 2s_t\cos\theta_t e_t^s - 2s_t\sin\theta_t e_t^\theta,\\
\hat{\lambda}^y_t & = \hat{\lambda}^y_{t - 1} + 2s_t  (1 + e_t^s)  \sin(\theta_t + e_t^\theta)\\
& \approx \hat{\lambda}^y_{t - 1} + 2s_t\sin\theta_t + 2s_t\sin\theta_t e_t^s + 2s_t\cos\theta_t e_t^\theta,\\
\end{aligned}
\end{equation}
where $\hat{\lambda}_t^x$ and $\hat{\lambda}_t^y$, in units of km, are the reported zonal and meridional displacements of a ship to its position since the last celestial correction.
$e_t^s$ and $e_t^\theta$ denote the errors in the relative ship speed and ship heading.
We assume that $e_t^s, e_t^\theta$ follow $N\left(0, \tau_s^2\right)$, $N\left(0, \tau_\theta^2\right)$, where $\tau_s, \tau_\theta$ calibrate the  magnitudes of uncertainties in the relative ship speed and ship heading, respectively. 
$e_t^s, e_t^\theta$ are assumed to be independent across time. For $t \in \mathcal{C}$, we have
\begin{equation}
\hat{\lambda}^x_t \sim N(0, (\tau_x \cos\psi_t)^2), \ \ \
\hat{\lambda}^y_t \sim N(0, \tau_y^2),
\end{equation}
where $\tau_x$ and $\tau_y$ calibrate the magnitudes of uncertainties in celestial correction. Combining linearized dead reckoning and celestial correction, we can approximate the variances of jumps in the longitudinal and latitudinal direction by
\begin{equation}
\begin{aligned}
\text{Var}(J^x) & =\tau^{2}_{s} \Delta x^{2} + \tau^{2}_\theta \Delta y^{2} + 2 \left(\tau_{x}  \cos\psi\right)^2,\\
\text{Var}(J^y) & =\tau^{2}_{s} \Delta y^{2} + \tau^{2}_\theta \Delta x^{2} + 2 \tau^2_{y}.
\end{aligned}
\label{eq:linearized_variance}
\end{equation}
$J^x, J^y$ denote the jumping distances in the longitudinal and latitudinal direction, and $\Delta x^2$ and $\Delta y^2$ denote the sum of squared distances between consecutive reports from the last celestial update to the position right before the next celestial update. We drop the dependence on $t$ for notational convenience.

We use $20,694$ midnight jumps identified from 943 HQ2 tracks to estimate the navigational parameters $\tau_x, \tau_y, \tau_s, \tau_\theta$. All the jumps are binned by 20km$\times$20km grids. Within each bin, we obtain the sample variances, $\widehat{V}_x$ and $\widehat{V}_y$, of the jumping distances in the longitudinal and latitudinal direction. Approximately the sample variances in each bin follow
\begin{equation}
\label{eq:variancelikelihood}
(n-1)\frac{\widehat{V}_x}{\text{Var}(J^x)}\sim\chi^2_{n-1}, \ \ \ (n-1)\frac{\widehat{V}_y}{\text{Var}(J^y)}\sim\chi^2_{n-1},
\end{equation}
where $n$ is the sample size in that bin. We set up standard non-informative priors on all the navigational parameters, and combine the likelihood specified in Equation \eqref{eq:variancelikelihood} to obtain the posterior distributions. 

The results are summarized in Table \ref{table:LinearResult}. For dead reckoning, due to the linearization, we obtain smaller estimates of the uncertainties in the relative ship speed and ship heading, compared to the results in Table  \ref{table:BayesHierarResult}. For celestial correction, we see that the two methods produce consistent estimates of the uncertainties.

\begin{table*}
\begin{minipage}{\textwidth}
\centering
\caption{Posterior distributions of the navigational parameters based on the linearized model.}
\begin{tabular}{|c|c|c|c|c|c|c|}
\hline
Quantiles & 5\% & 25\% & 50\% & 75\% & 95\%  & std\\
\hline
$\tau_{x}$\blfootnote{$\tau_x$ and $\tau_y$ denote the uncertainty in celestial correction along the longitudinal and the latitudinal direction, respectively. $\tau_s$ and $\tau_\theta$ denote the uncertainty in the relative ship speed and ship heading.} (km) & 30.91 & 31.12 & 31.27 & 31.44 & 31.67 & 0.23\\
\hline
$\tau_{y}$\ (km)& 24.17 & 24.37 & 24.50 & 24.64 & 24.85 & 0.21\\
\hline
$\tau_s$\hspace{0.12cm} (\%) & 15.07 & 15.25 & 15.38 & 15.49 & 15.65 &  0.17\\
\hline
$\tau_\theta$\ (rad) & 0.058 & 0.061 & 0.063 & 0.064 & 0.067 & 0.003\\
\hline
\end{tabular}
\label{table:LinearResult}
\end{minipage}
\end{table*}

\subsection{Additional checks}
\begin{sidewaysfigure}
\includegraphics[width=55pc]{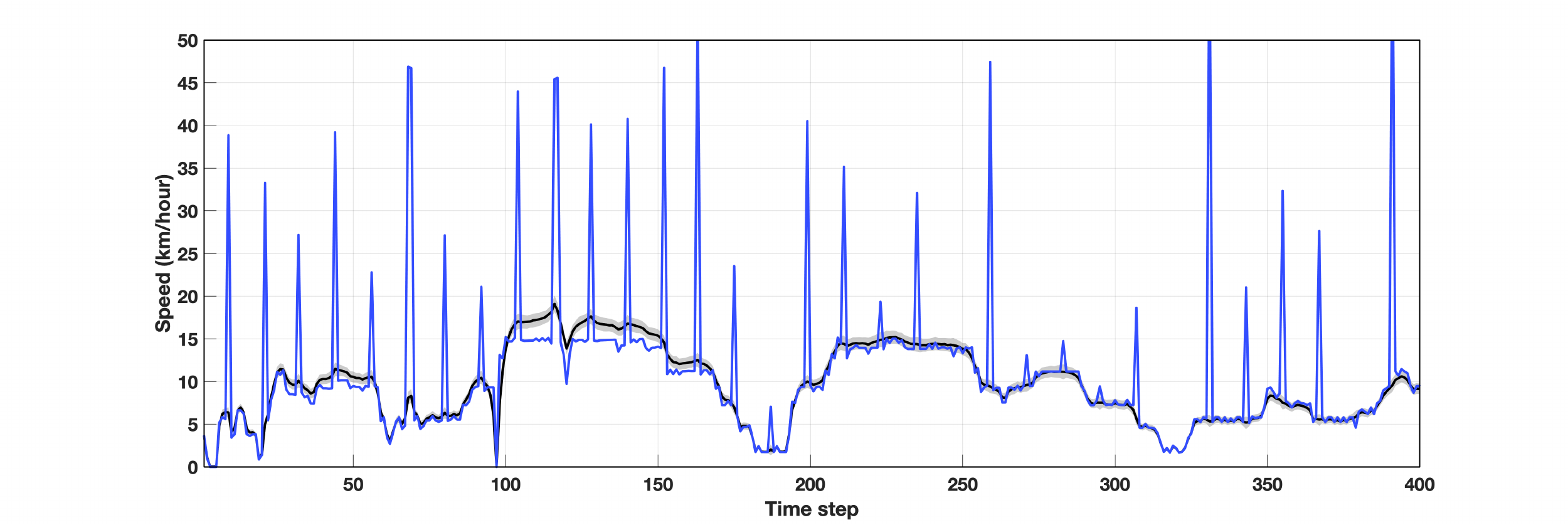}
\caption{The posterior distribution of ship speed $s_t$ for HQ2 track No.30. The blue line represents the empirical ship speed, the black line represents the posterior median, and the gray shadow calibrates the first and the third quantiles of the posterior distribution. The blue peaks correspond to the celestial corrections along the ship trajectory.}
\label{fig:speed}
\end{sidewaysfigure}

Figure \ref{fig:speed} shows the posterior distribution of ship speed $s_t$ for HQ2 track No.30. The blue line represents the empirical ship speed, the black line represents the posterior median, and the gray shadow calibrates the first and the third quantiles of the posterior distribution. We note that the unphysical blue peaks correspond to the large celestial updates along the ship trajectory, and the proposed state-space time series model in Section \ref{section:learnable 2-hourly} helps reasonably smooth out $s_t$.

\begin{figure*}
\centering
\includegraphics[width=14pc]{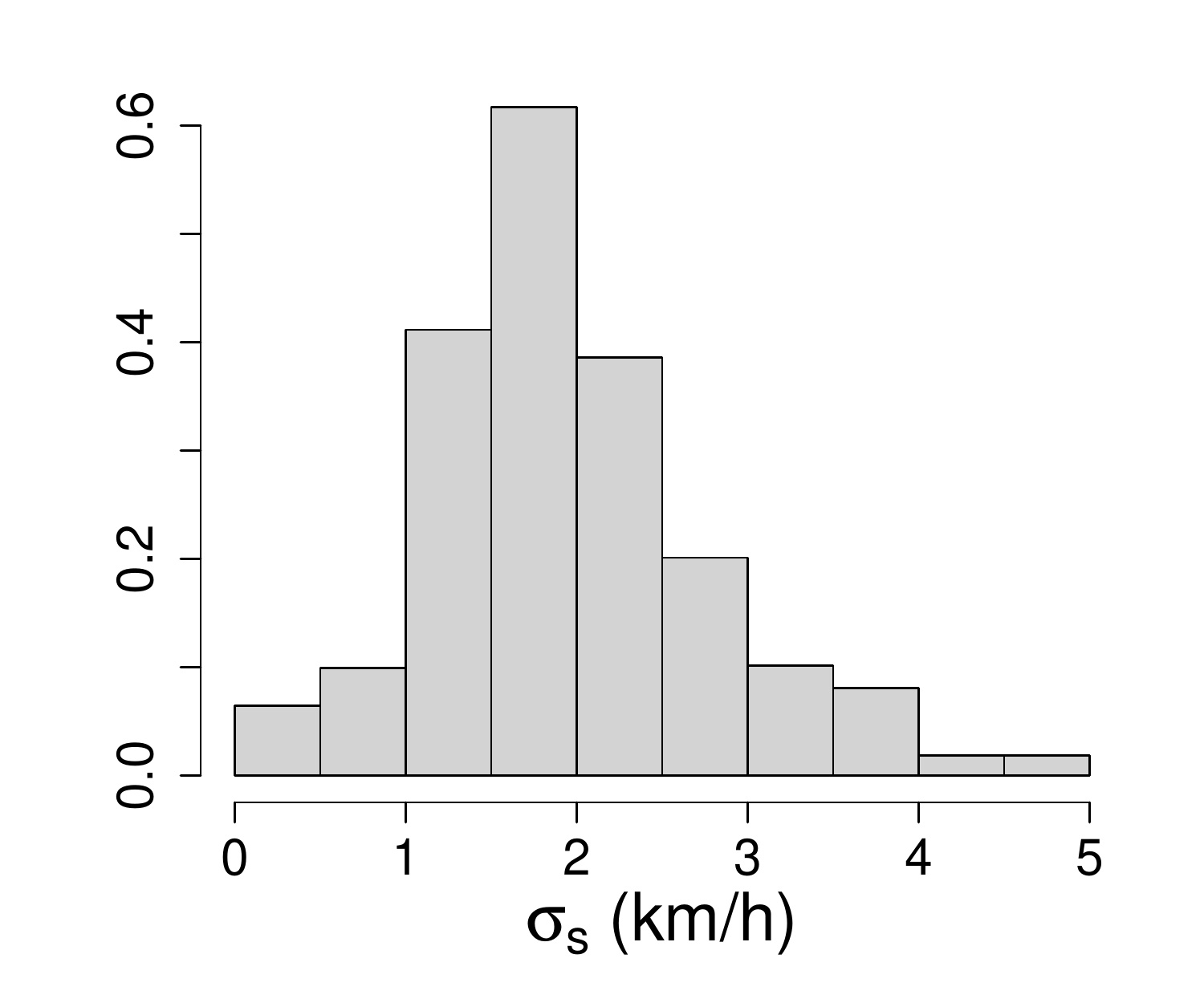}
\includegraphics[width=14pc]{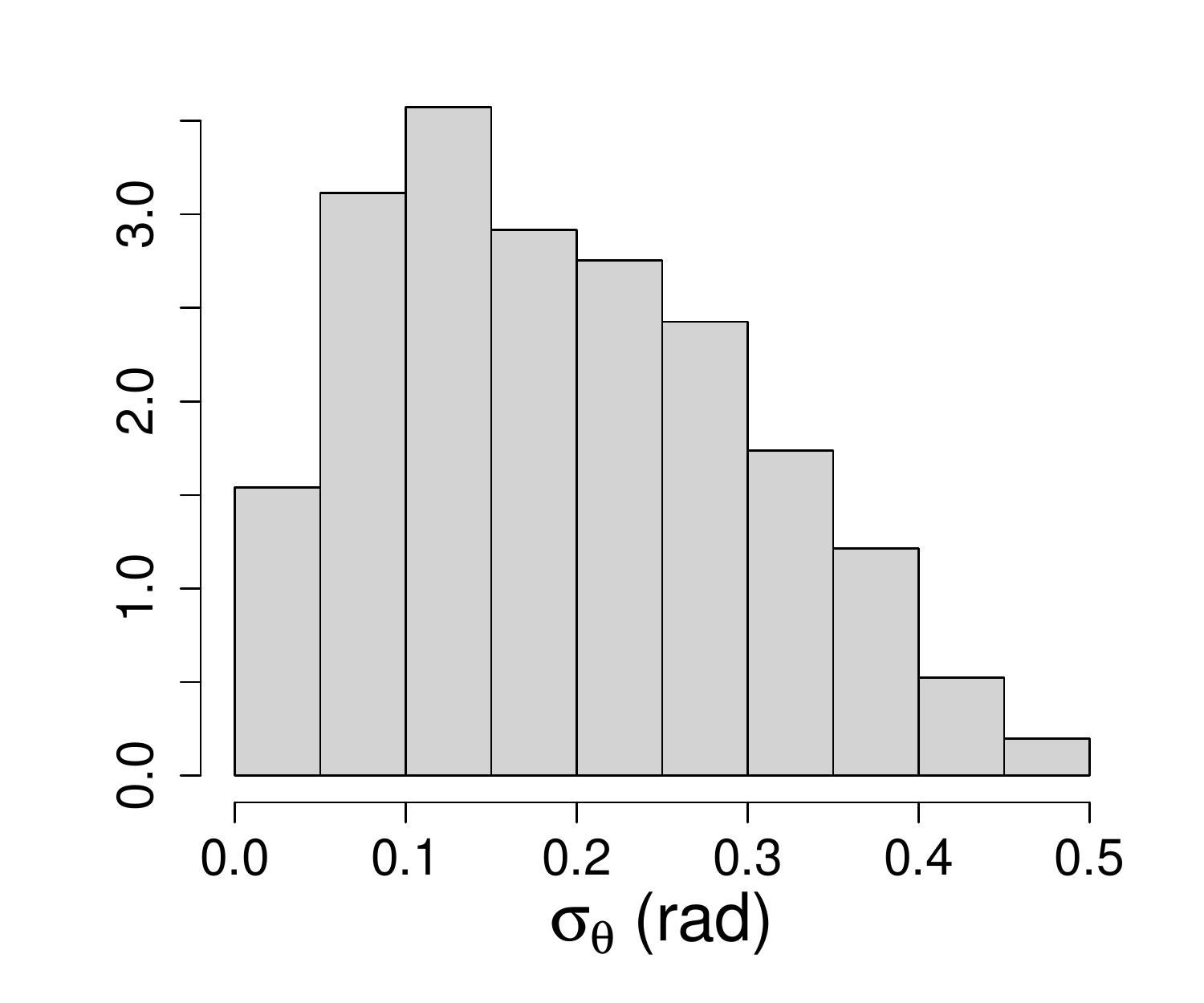}
\caption{The histograms of the posterior means of the navigational parameters $\sigma_s$ and $\sigma_\theta$ across HQ2 tracks, which calibrate the evolutionary uncertainties in  ship speed and heading, respectively. }
\label{fig:evolution-variance}
\end{figure*}

Figure \ref{fig:evolution-variance} demonstrates the distributions of the navigational parameters $\sigma_s$ and $\sigma_\theta$\footnote{See the definitions of $\sigma_s$ and $\sigma_\theta$ in Equations \eqref{eq:transition-sp}, \eqref{eq:transition-th} and the discussion therein.} across HQ2 tracks, which calibrate the evolutionary uncertainties in ship speed and heading, respectively.  We see that $\sigma_s$ is small enough so that the lower truncation on $\epsilon_t^s$ has little effect. In addition, although $\theta_t$ is a bounded quantity, $\sigma_\theta$ is also sufficiently small so that there are no unexpected
consequences to assume a normal distribution for $\epsilon_t^\theta$.

\begin{figure*}
\centering
\includegraphics[width=14pc]{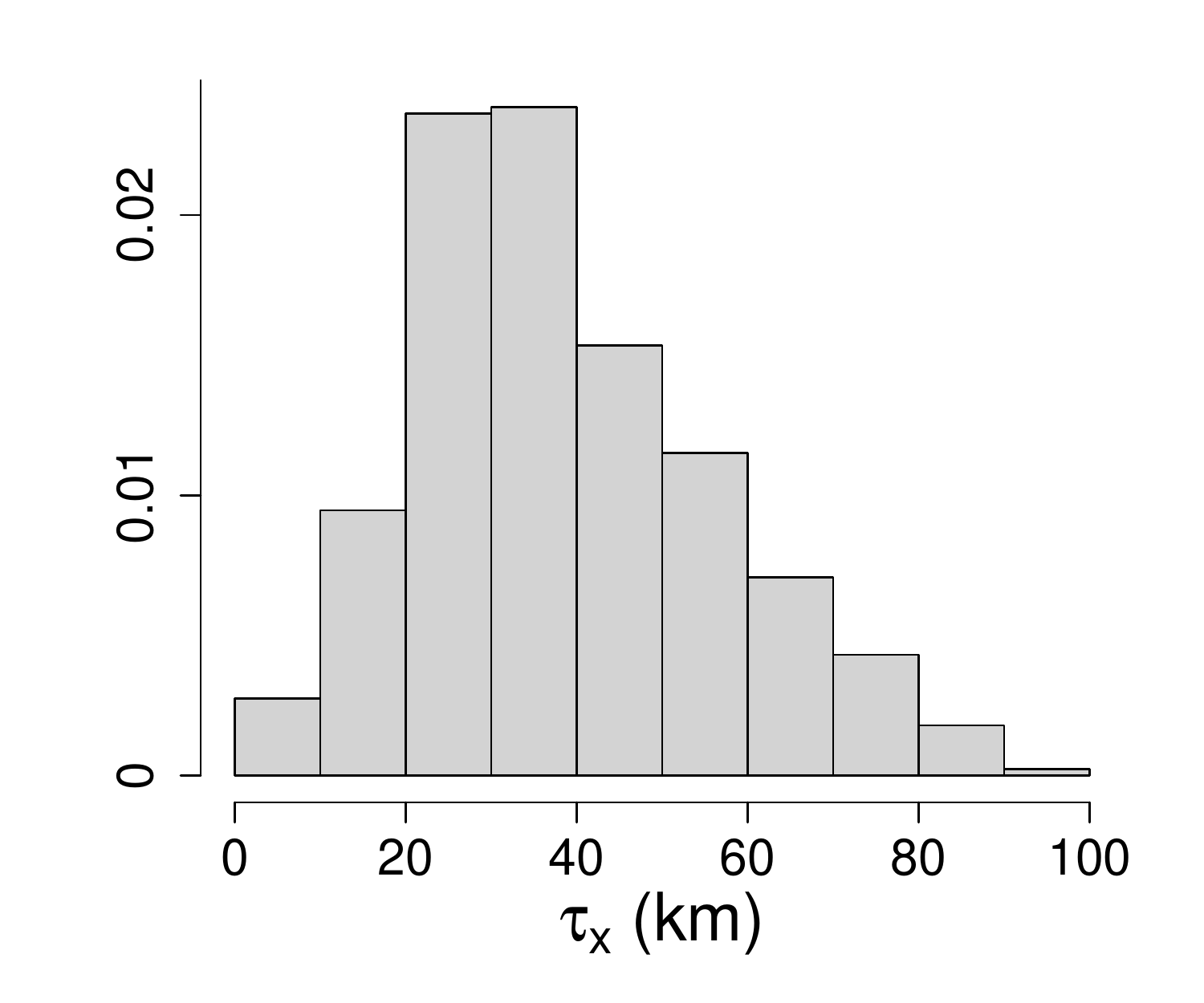}
\includegraphics[width=14pc]{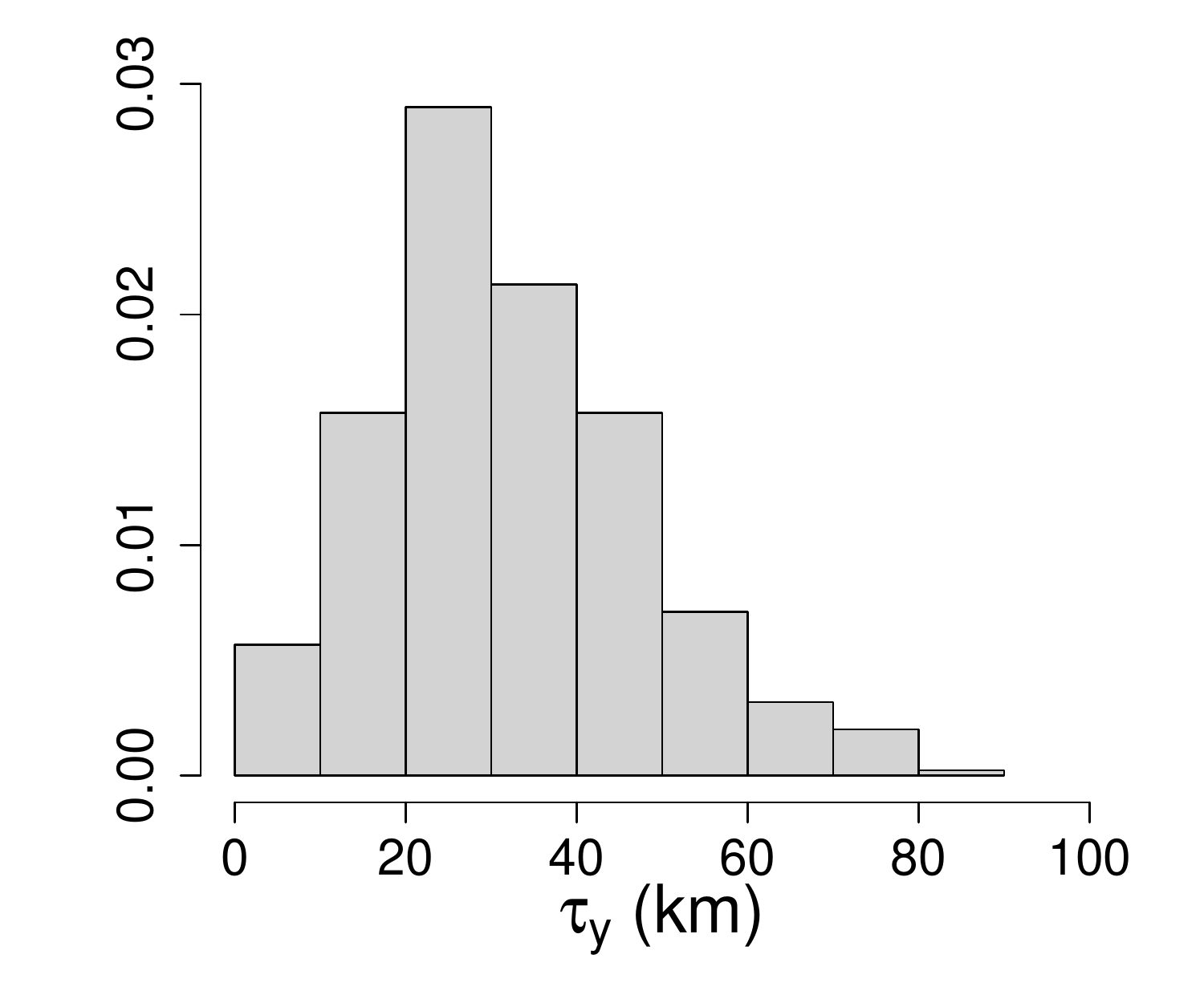}
\includegraphics[width=14pc]{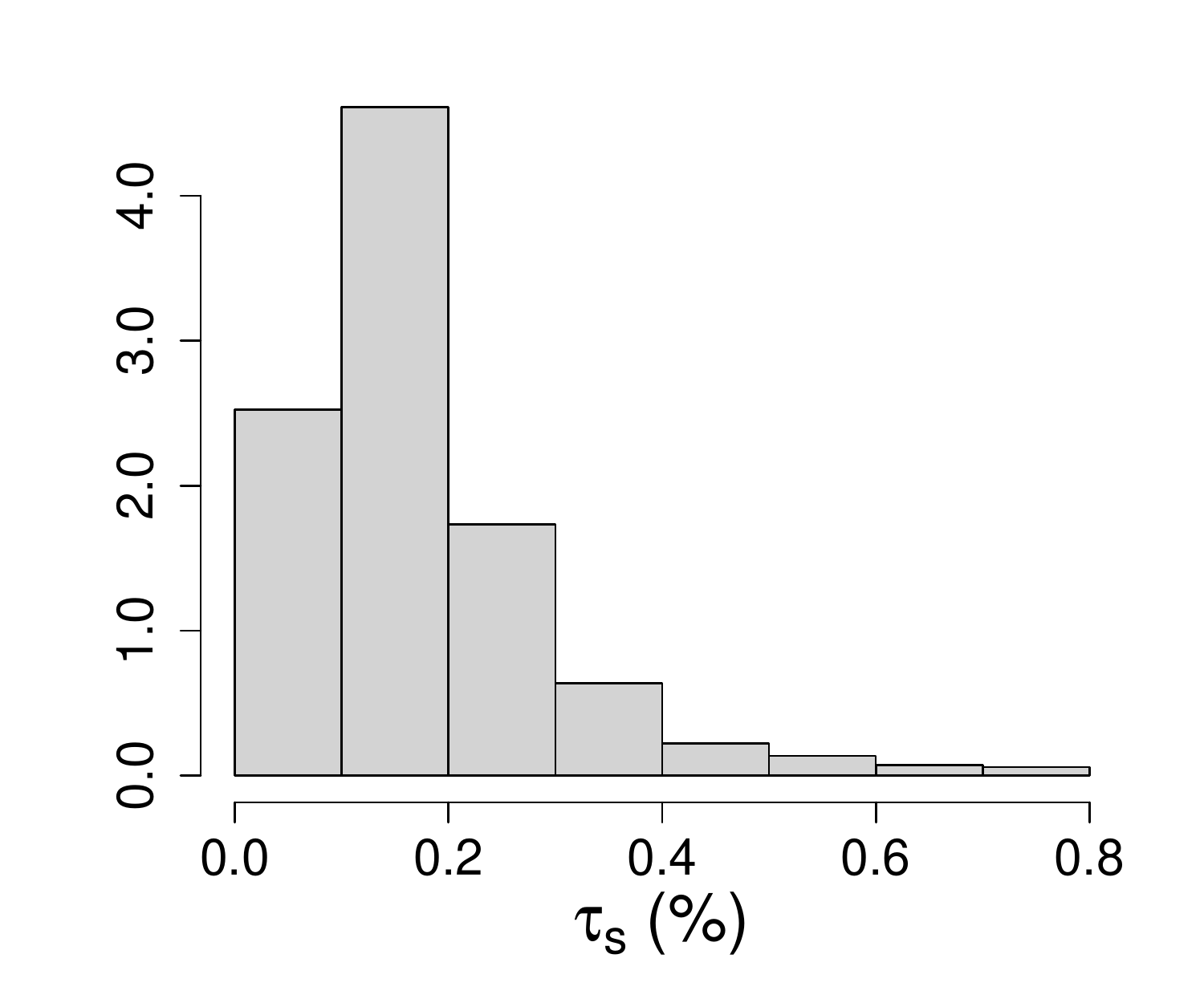}
\includegraphics[width=14pc]{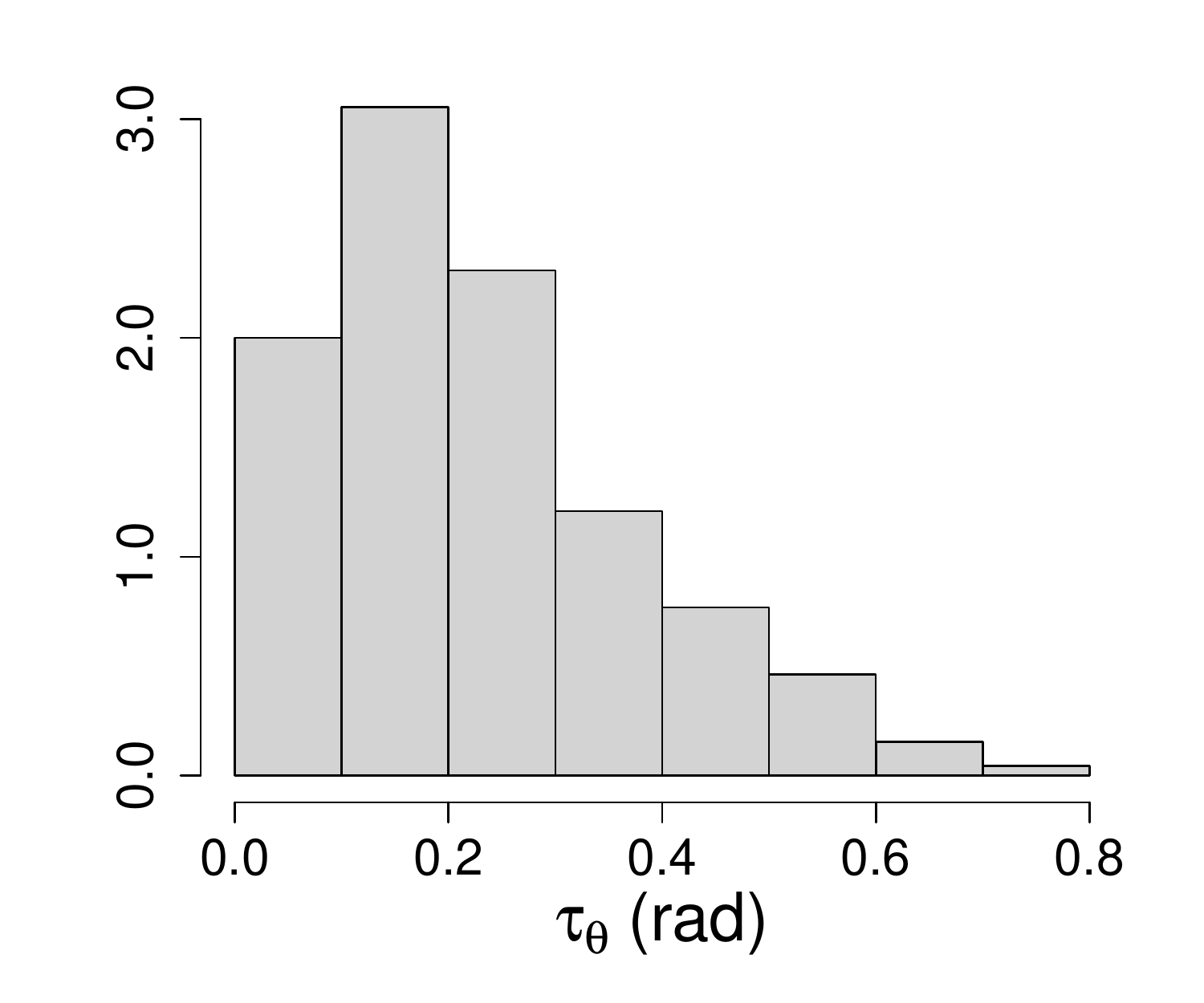}
\caption{The histograms of the posterior means of the navigational parameters across HQ2 tracks. $\tau_x$ and $\tau_y$ calibrate the uncertainties in celestial correction along the longitudinal and the latitudinal direction, respectively. $\tau_s$ and $\tau_\theta$ calibrate the uncertainties in the relative ship speed $\hat{s}_t/s_t$ and ship heading, respectively. }
\label{fig:variability}
\end{figure*}

We also summarize the variability of the navigational parameters across HQ2 tracks in Figure \ref{fig:variability}. In particular, we show the histograms of the posterior means of the navigational parameters $\tau_x$, $\tau_y$, $\tau_s$, and $\tau_\theta$ for HQ2 tracks. 
%%%%%%%%%%%%%%%%%%%%%%%%%%%%%%%%%%%%%%%%%%%%%%%%%%%%%%%%%%%%%%%%%%%%%%%%%%%%%%%%%%%%%%%%%%%%%%%%

\bibliographystyle{apalike}
\bibliography{SST.bib}

\end{document}